\documentclass{article}

\usepackage{acronym}
\usepackage{algorithmic}
\usepackage{amsmath}
\usepackage{amssymb}
\usepackage{color}
\usepackage{graphicx}
\usepackage{subfigure}
\usepackage{longtable}
\usepackage{url}
\usepackage{natbib}
\usepackage[section]{placeins}
\usepackage{times}
\usepackage{xcolor}
\usepackage[pdftex,colorlinks,citecolor=blue,urlcolor=blue,linkcolor=blue]{hyperref}
\usepackage{setspace}
\acrodef{ASK}[ASK]{Anomalous States of Knowledge}
\acrodef{ANOVA}[ANOVA]{Analysis of Variance}
\acrodef{API}[API]{Application Programming Language}
\acrodef{AIC}[AIC]{Akaike Information Criterion}
\acrodef{BIC}[BIC]{Bayes Information Criterion}
\acrodef{AP}[AP]{Average Precision}
\acrodef{BIR}[BIR]{Binary Independence Retrieval}
\acrodef{BM25}[BM25]{Best Match N. 25}
\acrodef{CFA}[CFA]{Confirmatory \ac{FA}}
\acrodef{EFA}[EFA]{Exploratory \ac{FA}}
\acrodef{FA}[FA]{Factor Analysis}
\acrodef{CFI}[CFI]{Comparative Fit Index}
\acrodef{TLI}[TLI]{Tucker-Lewis Index}
\acrodef{CG}[CG]{Cumulative Gain}
\acrodef{CNF}[CNF]{Conjunctive Normal Form}
\acrodef{CTR}[CTR]{click-through rate}
\acrodef{DCG}[DCG]{Discounted Cumulative Gain}
\acrodef{DNF}[DNF]{Disjunctive Normal Form}
\acrodef{GIR}[GIR]{Geometry of \ac{IR}}
\acrodef{HAL}[HAL]{Hyperspace Analogue to Language}
\acrodef{HITS}[HITS]{Hyperlinked Induced Topic Search}
\acrodef{KLD}[KLD]{Kullback-Liebler Divergence}
\acrodef{IDF}[IDF]{Inverse Document Frequency}
\acrodef{IN}[IN]{Information Need}
\acrodef{IRF}[IRF]{Implicit \acl{RF}}
\acrodef{IRS}[IRS]{\ac{IR} System}
\acrodef{IR}[IR]{Information Retrieval}
\acrodef{IIR}[IIR]{Interactive Information Retrieval}
\acrodef{LETOR}[LETOR]{Learning To Rank}
\acrodef{LM}[LM]{Language Model}
\acrodef{LSA}[LSA]{Latent Semantic Analysis} 
\acrodef{MAP}[MAP]{Mean \acl{AP}}
\acrodef{MLE}[MLE]{Maximum Likelihood Estimation}
\acrodef{ML}[ML]{Machine Learning}
\acrodef{MLM}[MLM]{Multilevel Modeling}
\acrodef{MR}[MR]{Multiple Regression}
\acrodef{MSE}[MSE]{Mean Squared Error}
\acrodef{MSLR}[MSLR]{Microsoft Learning-to-Rank}
\acrodef{NDCG}[NDCG]{Normalized Discounted Cumulative Gain}
\acrodef{NIST}[NIST]{{National Institute of Standard and Technology}}
\acrodef{NPL}[NPL]{Neyman-Pearson Lemma}
\acrodef{PBS}[PBS]{Polarizing Beam Splitter}
\acrodef{PA}[PA]{Path Analysis}
\acrodef{QPP}[QPP]{Query Performance Prediction}
\acrodef{PRF}[PRF]{Pseudo Relevance Feedback}
\acrodef{PRP}[PRP]{Probability Ranking Principle}
\acrodef{P@r}[P@$r$]{Precision at rank $r$}
\acrodef{QD}[QD]{Quantum Detection}
\acrodef{QIRB}[QIRB]{Quantum Information Retrieval Basis}
\acrodef{QLM}[QLM]{Query Language Model}
\acrodef{QLRA}[QLRA]{Quantum-Like Representation Algorithm}
\acrodef{QM}[QM]{Quantum Mechanics}
\acrodef{RAM}[RAM]{Reticular Action Model}
\acrodef{RF}[RF]{Relevance Feedback}
\acrodef{RMSEA}[RMSEA]{Root Mean Square Error of Approximation}
\acrodef{SRMR}[SRMR]{Standardized Root Mean Square Residual}
\acrodef{RSJ}[RSJ]{Robertson and Sparck-Jones}
\acrodef{SC}[contextual variable]{Contextual Variable}
\acrodef{SEM}[SEM]{Structural Equation Modelling}
\acrodef{SERP}[serp]{search engine result page}
\acrodef{SVD}[SVD]{Singular Value Decomposition} 
\acrodef{TFIDF}[TFIDF]{\ac{TF} $\times$ \ac{IDF}}
\acrodef{TF}[TF]{Term Frequency}
\acrodef{TREC}[TREC]{{Text REtrieval Conference}}
\acrodef{TRW}[TRW]{Term Relevance Weight}
\acrodef{URL}[URL]{Uniform Resource Locator}
\acrodef{VSM}[VSM]{Vector Space Model}
\acrodef{WWW}[WWW]{World Wide Web}
\acrodef{iid}[i.i.d.]{independent and identically distributed}
\acrodef{qrel}[qrel]{relevance assessment}
\acrodef{RIA}[RIA]{Reliable Information Access}

\newcommand{\var}[1]{\text{var}(#1)}

\newcommand{\cor}[2]{\text{cor}(#1,#2)}

\newcommand{\papertitle}{Evaluation of Information Retrieval Systems Using Structural Equation Modelling}

{}{}
\newcommand{\citeN}{\cite}
\newcommand{\citeANP}{\cite}

\title{\papertitle}
\author{Massimo Melucci}
\begin{document}
\maketitle
\begin{abstract}
  The interpretation of the experimental data collected by testing systems across input datasets and model parameters is of strategic importance for system design and implementation. In particular, finding relationships between variables and detecting the latent variables affecting retrieval performance can provide designers, engineers and experimenters with useful if not necessary information about how a system is performing. This paper discusses the use of \ac{SEM} in providing an in-depth explanation of evaluation results and an explanation of failures and successes of a system; in particular, we focus on the case of \acl{IR}.
\end{abstract}
\newpage
\acresetall
\tableofcontents
\newpage
\acresetall

\section{Introduction}
\label{sec:intro}

Humans often have to find solutions to problems. The attempts to find solutions
are the main causes of information needs. To meet information needs, users
search for relevant information while avoiding useless ones.  The aforementioned
context is where \ac{IR} systems perform the complex of activities to represent
and retrieve documents containing information relevant to user's information
needs, thus becoming a crucial function of computerised information systems.

Effective retrieval systems should be designed to obtain high
precision\footnote{The proportion of retrieved documents that are found
  relevant.} and high recall\footnote{The proportion of relevant documents that
  are retrieved.}.  To obtain a measure of retrieval effectiveness, designers
and experimenters employ a variety of test collections, since the effectiveness
of a retrieval system may widely vary according to queries and retrieval
algorithms; for example, \cite{Harman&09} report that large variations in measures
of effectiveness may be observed for \ac{RF} when varying the number of feedback
documents and terms.

Understanding the reasons of retrieval failures and measuring the room for
effectiveness improvement is of strategic importance for system design and
implementation.  The interpretation of the experimental data collected by
testing retrieval systems across variables would help designers and researchers
to explain whether and when a system or a component thereof performed better or
not than another system or component.  

Despite the unquestionable importance of in-depth analysis of experimental
results, many research papers fail to provide insights into experiments, apart
from some statistical significance tests which, however, rarely point out
retrieval model weaknesses.  One reason for the lack of methodologies supporting
researchers and experimenters in interpreting the retrieval failures is the
absence of a language that can help communicate in spoken or written words,
variables and causal relationships thereof.

The principal purpose of this paper is thus to explain how to fill the gap
between a mere -- even though necessary -- description of tables, graphs and
statistical testing, on the one hand, and the use of advanced statistical
methods to describe the variables and their relationships that characterise
retrieval performance in a more natural way than traditional statistics. We
argue that \ac{SEM} can be such a methodology.

The paper is structured as follows. Section \ref{sec:related} describes the
context of the paper and mentions some relevant related work.  Section
\ref{sec:remarks} remarks on the use of \ac{SEM} in \ac{IR} and explains the
main differences among analysis methods.  In Section \ref{sec:uses}, we explain
how \ac{SEM} can be applied to \ac{IR} by means of a series of experimental case
studies.  Section \ref{sec:conclusions} comments on the potentiality of \ac{SEM}
in \ac{IR}.

\section{Related Work}
\label{sec:related}

\ac{SEM} is a general methodology encompassing multivariate methods addressed in
\ac{IR} since \citeN{Salton79}'s research work; other notable examples include
\citeN{Deerwester&90}'s \ac{LSA} and other \ac{FA} methods utilised in
contextual search \citep{Melucci12b}.  

The \ac{IR} community has already developed some approaches to analyzing the
causes of both missing relevant documents and the retrieval of irrelevant
documents; reliability analysis, retrievability analysis, query performance
prediction and axiomatic analysis are the most utilised to this end.  Some
approaches might have been missed, however, those mentioned are the principal
approaches in our opinion and to our knowledge.

\subsection{Reliability, Retrievability, \acl{QPP} and Rank Correlation}
\label{sec:reli-retr-qpp}

\subsubsection{Reliability}
\label{sec:reliability}

{Reliability} is concerned with situations where a system retrieves relevant
documents and misses non-relevant documents across a set of queries. A major
factor in the unreliability of a system is the extremely large variation in
performance across queries.  When different systems or variants are considered,
variation can also be caused by system algorithms and implementations.

A systematic approach to understanding the reasons why systems fail in
retrieving relevant documents or succeed in retrieving irrelevant documents has
been implemented by the \ac{RIA} workshop documented by \citeN{Harman&09}.  We
summarise the main outcomes as follows:
\begin{itemize}
\item although systems tend to retrieve different document sets, they tend to
  fail for the same reason, i.e. wrong query understanding due to, for example,
  over/under stemming or missed synonyms;
\item systems not only tend to emphasize the same query aspects, but they also
  emphasize wrong aspects;
\item \citeN{Buckley09} reported that variations in system performance can occur
  \begin{itemize}
  \item across queries in terms of \ac{AP}, thus calling for an analysis at the
    level of query, and
  \item across systems or variants thereof, e.g. particular devices such as
    relevance feedback or query expansion;
  \end{itemize}
\item most of the average increase of effectiveness of query expansion is due to
  a few queries that are greatly improved;
\item performance is increased by several good terms and cannot be increased by
  one single crucial term;
\item along these lines, \citeN{Ogilvie&09} suggested cross-validation to find
  the best number of terms.
\end{itemize}
Approaches inspired to data mining to understanding retrieval failures were also
proposed by \citeN{Bigot&11}. Reliability analysis also investigated the best
practices for learning to rank deployments by \citeN{Macdonald&13}.  The
analysis reported was performed starting from a series of research hypotheses
about the impact of sample size, type of information need, document
representation, learning to rank technique, evaluation measure, and rank cutoff
of the evaluation measure on the observed effectiveness.  The methodology that
was implemented by \citeANP{Macdonald&13} to perform the analysis was based on
the definition of some variables and three research themes, i.e. sample size,
learning measure and cutoff, learning cutoff and sample size; the research
themes were associated to the variables, which were labeled as either fixed or
factor. Sample size definition was also addressed by \citeN{Voorhees&02} using
empirical error rates, as well as by \citeN{Sakai14} using power analysis,
paired t-test, and \ac{ANOVA}.  Moreover, \citeN{Bailey&15} too reported that
the system performance variations of a single system across queries is
comparable or greater than the variation across systems for a single query.

\subsubsection{Retrievability}
\label{sec:rertievability}

{Retrievability} concerns the variations between systems with respect to the
rank of the same retrieved document, according to \citeN{Azzopardi&08}.
Retrievability may also depend on the subsystems (e.g. crawlers) that decide
which documents are indexed, the way users formulate queries, the retrieval
functions, the user's willingness to browse document lists, and the system's
user interface.  Many systems make many documents little retrievable and rank
documents in lists that would not change were little retrievable documents
removed from the index.  A measure of retrievability of document $d$ was
proposed in \cite{Azzopardi&08}:
\begin{equation}
  \label{eq:retrievability}
  \mbox{ret}(d) = \sum_{q \in Q} L(q) f(r(d,q), r^*)
\end{equation}
where $Q$ is set of queries, $r(d,q)$ is the rank of $d$ in the retrieved
document list, $L(q)$ is the likelihood of $q$, $r^*$ is the maximum examined
document rank, and $f$ is the cost/utility of $d$.  The computation of
$\mbox{ret}(d)$ is challenging since it should be estimated across many
different systems and many different queries.  Low retrievability causes
retrieval bias since a system may favour the most retrievable documents.
\citeN{Wilkie&14} reported that a negative correlation exists between retrieval
bias and some retrieval performance measures, thus suggesting that reducing
retrieval bias would increase performance.

\subsubsection{\acl{QPP}}
\label{sec:qpp}

{\ac{QPP}} deals with situations where a \emph{specific} query fails or succeeds
in retrieving relevant documents, whereas retrievability analysis is only based
on using document features and reliability analysis is based on query sets.  A
measure of query ambiguity and then of a \ac{QPP} called query clarity was
proposed by \citeN{Cronen-Townsend&02a} and further improved and extended by
\citeN{Hauff&08}.  The intuition behind query clarity is that, the more
different the query language from the collection language, the less the
ambiguity and then the better the retrieval performance.  The clarity score of a
query is the \ac{KLD} between the collection language and the query language.
The query language is estimated by the set of retrieved documents matching the
query.  The more diverse the latter and the more similar it is to the collection
language, the more the query is ambiguous.  \ac{QPP} usually estimates
effectiveness without relevance judgments, but using retrieved document
features. However, assessing very few top-ranked documents can dramatically
improve \ac{QPP} quality according to \citeN{Butman&13}.
\citeN{Zhao&08,Zhou&06b} proposed further measures and techniques.  Moreover,
\citeN{Hauff&10a} found that the user's predictions of query performance do not
correlate with the system's predictions; on the other hand, different approaches
were described by \citeN{Kurland&12} in one uniform framework; association rules
were applied to the discovery of poorly performing queries by \citeN{Kim&13};
and some explanations of why \ac{QPP} might not work as expected were reported
by \citeN{Raiber&14}.  \citeN{Cummins14} proposed to predict query performance
from document score distributions and also provides a good and up-to-date survey
of \ac{QPP}.

\subsubsection{Rank Correlation}
\label{sec:rank-correlation}

An alternative approach to comparing runs might be based on rank correlation
measurement.  Rank correlation refers to a family of statistical measures of the
degree to which two rankings should be considered similar, that is, the items of
a ranking are disposed approximately in the same order as the same items in
another ranking; examples of rank correlation measures are the $\tau$
coefficient by \cite{Kendall38} and the $\rho$ coefficient by
\cite{Spearman&04}. 

The main advantage of rank correlation measures is the simplicity of measuring
the degree to which two rankings are similar using one single number, which may
be tested for significance because it can often be provided with a probability
distribution under the null hypothesis of incorrelation when samples are large
enough.  

The main weakness of rank correlation measures is the poor description
capability, because these measures are unable to distinguish between exogenous
variables and endogenous variables and between latent and manifest variables.  A
rank correlation measure is a zero-dimensional measure whereas a structural
equation model is a multidimensional measure; for example, if Kendall's tau of
the correlation between two measures of effectiveness may be statistically
significant, but if the value is, say, 0.485, the coefficient is little
informative about the differences between the tested systems.

\subsubsection{Comparison to \ac{SEM}}
\label{sec:comparison-sem}

Retrievability, query ambiguity and \ac{QPP} are related each other.
Retrievability depends on query ambiguity, since an ambiguous query is more
likely to select less relevant documents than an unambiguous query.  Moreover,
\ac{QPP} is obviously related to query ambiguity. Incorporating user variability
in system-based evaluation is also somehow related to \ac{QPP}. User variability
allows the researchers to more precisely measure the effectiveness of the system
to different segments of the user base, thus allowing them to predict which
systems will be the most effective in performing a certain user's task; see the
papers by \citeauthor{Carterette&11}
\citeyear{Carterette11,Carterette&12}.

Reliability analysis, retrievability analysis, and \ac{QPP} are performed with
the idea that a retrieval system can be viewed as a black box in which
independent variables can be entered and dependent variables can be observed.
Following this idea, the variations of the latter can be explained by the
variations of the former.  Besides, this idea entails that retrieval systems are
indeed ``black boxes'' about which nothing can be known but that can be observed
when something is given to them as input -- the boxes' internal mechanisms are
hidden to the external observer.

A quite different approach to understanding retrieval failures and successes --
it might be named axiomatic -- was suggested by \citeN{Fang&04} and
\citeN{Fang&05}.  The basic idea of the axiomatic approach is that (1) some
heuristic rules can be defined to describe an effective retrieval function and
(2) the inefficacy of a retrieval function is related to the retrieval
function's failure to comply with these heuristic rules in the sense that the
rules are necessary conditions of effective retrieval, that is, the violation of
a rule determines a loss of effectiveness. The potential of the axiomatic
approach can be exploited to improve the retrieval functions violating the rules
as reported by \citeN{Fang&11}.

On the one hand, reliability analysis, retrievability analysis, and \ac{QPP} are
specific to \ac{IR}.  On the other hand, \ac{SEM} was investigated and applied
to complex social, economic, and psychological phenomena. For example,
attitudes, personality traits, health status, and political trends are often
variables of interest to sociologists.  Intellectual abilities of students or
teaching styles of instructors are important variables in education. The
relationship between demand and supply is very important to economists; some
examples are reported in Section \ref{sec:sem-other}.

\subsection{\acl{SEM} in \acl{IIR}}
\label{sec:sem-inter-ir}

\ac{SEM} is still in its infancy within laboratory-based \ac{IR}; in contrast,
it recently received a great deal of attention in \ac{IIR} because it provides
an effective framework to modeling/ the complex variables emerging from the
interaction between user and \ac{IR} system.  The theme of interaction between
user and system was at the root of \ac{IR} since the early Eighties when
\cite{Belkin&82a,Belkin&82b} addressed the problem of the \ac{ASK} as well as
\cite{Marchionini&88} and \cite{Marchionini&94} investigated how hypertext
systems can induce a novel approach to searching for information.

The occurrence of latent factors in the user's mind such as search task and
intent and the inherent difficulty in measuring these factors were the main
reasons why quantitative methods measuring latent factors by means of manifest
variables were suggested to assess the importance and the relationships among
variables and factors; to this respect, \ac{SEM} represents the most general
framework.  Therefore, in \ac{IIR}, \ac{SEM} has been drawing attention to a
degree that some tutorials such that that presented by \cite{Kattenbeck&18} are
becoming necessary or useful for systematizing the corpora of research articles
such as those authored by \cite{Zhang&14} and \cite{Ishita&17}.

In this paper, we limit ourselves to the use of \ac{SEM} in laboratory-based
\ac{IR} evaluation, which has received a little deal of attention, without
further addressing the already covered use in \ac{IIR}.

\subsection{Comparison to Other Approaches to Analyzing Experimental Data}
\label{sec:sem-other-analysis}

The statistical inference performed using experimental data provides some
guidance to see whether two systems performed to a similar degree; for example,
it helps decide whether the average difference in precision between system (or
component) performances is due to chance or it signals a diversity between the
systems (or components).  A statistical estimator measures the difference; the
p-value\footnote{The p-value of an observation is the probability of measuring a
  value greater than the absolute value of the observation when the observed
  values are expected to be zero (null hypothesis).  The p-value is then an
  indirect way to measure how far the observation is from the null hypothesis.}
of the estimator can measure the statistical significance of the estimated
value, that is, the degree to which the value should not be considered a random
fluctuation.  This approach to evaluating systems is indeed the standard
practice of evaluation as reported by many research papers.

Other questions about the reasons that a system or component performed better or
worse than another system or component would require further statistical
methodologies which are sadly less frequently reported in the literature on
evaluation. Indeed, an inferential analysis whether a system performed
differently from another is unable to explain retrieval performance
variations. A consequence of the lack of explanation of the differences in
performance between systems is the difficulty in improving retrieval performance
-- the retrieval performance observed for some queries can be improved when the
reasons that make the retrieval system ineffective become known to the
researchers.

\ac{SEM} may support researchers because it provide them with a language to
describe observed data. The opportunity -- and the necessity -- of choosing an
appropriate model is toward stimulating and helping researchers to explain their
experimental results beyond a mere -- even though necessary -- textual
description of tables, graphs and statistical testing. The dependency on the
experimenter's knowledge of the domain to which the process is applied (e.g.,
\ac{IR} experimentation) is a strength, since it makes the experimenter's point
of view explicit and reproducible.

\ac{SEM} can be viewed as generalization of other multivariate analysis.  In
this section we provide a comparison to help readers to understand the \ac{SEM}
advantages.  To this end, we prepared Table \ref{tab:comparison}.

\begin{table}[t]
  \centering
  \begin{tabular}[y]{|c|c|c|c|c|c|} \hline \multicolumn{1}{|r|}{Analysis}&
    Correlation & Regression & Path & Factor & \ac{SEM}\\ {Property}&
    Analysis &
    Analysis & Analysis & Analysis & \\ \hline Association & Y & Y & Y & Y & Y \\
    Directionality & N & Y & Y & Y & Y \\ Prediction & N & Y & Y & Y & Y \\
    Heterogeneity & N & N & Y & Y & Y \\ Latent Variables & N & N & N & Y & Y \\
    Latent Association & N & N & N & N & Y \\ Causality & N & N & N & N & Y \\
    \hline
  \end{tabular}
  \caption{For each column and row, 'Y' means that an analysis method
    (column) owns a property (row).  \emph{Association} means that two variables
    increase or decrease together; pure association means that association
    depends only on $X$ and $Y$, otherwise association is spurious.
    \emph{Directionality} means that a variable can be either exogeneous or
    endogeneous, and the influence of $X$ on $Y$ differs from the influence
    of $Y$ on $X$. \emph{Prediction} means that some independent
    variables determine, i.e. predict, some dependent variables.   \emph{Heterogeneity} means that some variables can be both
    exogeneous and endogeneous.  \emph{Latent variables} means that experimenters
    can define latent variables.  \emph{Latent association} means that
    experimenters can define association between latent variables.
    \emph{Causality} means that experimenters can test whether the hypothesis that
    one variable depends on another variable is confirmed by the observed
    data.}
  \label{tab:comparison}
\end{table}

Association is owned by every analysis method because mean and covariance --
i.e.  correlation -- is at the basis of more complex analysis.  If only
correlation matrices are used, correlation analysis cannot in its own
distinguish the direction of association.  Heterogeneity cannot even more so be
distinguished because covariance is commutative.  If there are three or more
variables, pure association can be measured using correlation, however, the
semantics of purity would make sense only if directionality held.  Heterogeneity
would imply that some variables determine other variables and therefore that
directionality holds.  Correlation between latent variables -- and association
thereof -- can only be estimated by manifest variables.

Regression extends correlation in that variables can be either exogeneous or
endogeneous because of directionality. (Regression coefficients are not
commutative.)  Regression detects pure association, since beta coefficients can
be calculated between variable pairs without the influence of third
variables. Path analysis can be represented by -- or is a specialisation of --
regression, since a variable can be both endogeneous (i.e. predicted) and
exogeneous.  Indeed, heterogeneity implies directionality.

\acf{FA} allows experimenters to extract latent factors in a partially
supervised manner, since an experimenter can extract some factors, but (1) s/he
cannot name them because their semantics is unknown until they are computed and
(2) s/he cannot model relationships between factors.  Indeed, factors are by
definition uncorrelated, yet rotation algorithms can rotate orthogonal factors
to obtain mutually oblique lines in a vector space.  \ac{FA} can be either
exploratory -- the number of factors is unknown -- or confirmatory -- mainly
concerned with testing hypotheses about the number of factors and the
significance of the relationships between factors and manifest variables.

The basic difference from \ac{SEM} is mainly concerned with estimating
relationships between latent variables, i.e. factors, whereas confirmatory
factor analysis is mainly concerned with the degree to which a factor determines
a manifest variable. As confirmatory factor analysis does not model supervised
association between latent variables as \ac{SEM} does, relationships between
factors can be indirectly represented by rotation only.

As for causality, it must be understood that \ac{SEM} does not discover causal
relationships between variables. \cite{Bollen&13} stated that ``researchers do
not derive causal relations from a [structural equation model]. Rather the
[structural equation model] incorporates the causal assumptions of the
researcher. These assumptions derive from the research design, prior studies,
scientific knowledge, logical arguments, temporal priorities, and other evidence
that the researcher can marshal in support of them. The credibility of the
[structural equation model] depends on the credibility of the causal assumptions
in each application.'' What \ac{SEM} can do is test the consistency between data
and variables which may be connected by causal relationships assumed by
researchers.

\ac{MLM} is quite related to \ac{SEM}, since it groups data into larger clusters
so that scores within each cluster may not be independent.  Recently,
\citeN{Crescenzi&16} have investigated \ac{MLM} to evaluate a number of
hypotheses about the effects of time constraint, system delays and user
experience.  \ac{MLM} and \ac{SEM} might converge to a single framework
according to \citeN{Kline15} and \citeN{Bartholomew&08}.

Besides \ac{SEM}, stepwise regression also selects the best predictors based on
statistical significance (i.e. p-value).  In practice, the predictor
showing the lowest p-value of its regression coefficient is selected and added
to the model.  After the addition of the best predictor, the worst predictors
showing the highest p-values or the p-values above a threshold are removed from
the model.  Although a stepwise regression function may compute the best model
in a short time, automatic predictor selection may depend on the solution of the
actual sample utilised to fit the model while another sample might suggest
another model \citep{Kline15}.

\subsection{\acl{SEM} and Other Domains}
\label{sec:sem-other}

In addition to the investigation of socio-economic phenomena, some uses of
\ac{SEM} regarded research areas that are somehow relevant to \ac{IR}, since the
factors affecting users' access to information systems were investigated.
\ac{SEM} was utilised by \citeN{Chan&05} to examine the multiple causal
relationships among the performances for different tasks (modeling, query
writing, query comprehension) performed by the users of a database interface, in
which the data model and query language are major components.  A structural
equation model was also used to investigate users' behaviour within community
networks\footnote{``[G]eographically based Internet services that provide local
  residents a full range of Internet services and other information and
  communication technology related services, including computer and Internet
  training, setting up public access sites, the creation of digitised local
  information database, and organisational ICT consulting.''  \cite{Kwon&05}} by
\citeN{Kwon&05}, Bulletin Board Systems by \citeN{Chen&07}, Wikipedia by
\citeN{Cho&10}, social network systems by \citeN{Kipp&10} and \citeN{Park&14},
electronic commerce by \citeN{Lu&10a} and \citeN{Afzal&13}, library systems by
\citeN{Sin&10}, exploratory search by \citeN{OBrien&13}, agile software
development by \citeN{Senapathi&14}, and online education by \citeN{Zhang&15}.
\citeN{Kher&09} used a variation of \ac{SEM} called Latent Growth Modeling to
study longitudinal data where time is a relevant variable.

\section{Remarks on the Use of \acl{SEM}}
\label{sec:remarks}

The basic idea underlying the use of \ac{SEM} in \ac{IR} is that the evaluation
of indexing and retrieval of large and heterogeneous document collections
performed by an \ac{IR} system may be viewed as a phenomenon similar to the
social and economic phenomena investigated by \ac{SEM}.  According to this view,
an investigator is supposed to be unable to explain all the reasons why a system
failed or succeeded in performing indexing and retrieval operations, since the
complexity of the document collections and of the user's queries can be at the
level that goes beyond the potential of the investigator's instruments.  The
complexity might not be caused by the retrieval system's software architecture
-- it can be well known and documented -- rather, it may be due to the
heterogeneity of the document collection and the context-sensitiveness of the
user's interaction and relevance assessment.

However complex the evaluation of indexing and retrieval of large and
heterogeneous document collections may be, our rationale is that \ac{IR}
evaluation results can be described by causal hypotheses between variables using
\ac{SEM}, where the variables are both latent or manifest quantities that are
taken as input while regression coefficients, beta coefficients and fit indexes
are given as output.

This section illustrates the main properties that make \ac{SEM} suitable for
\ac{IR} experimental results investigation.  In summary, we will explain the
following reasons: experiments consists of observing manifest variables;
covariation (e.g. between relevance assessments and frequency) is at the basis
of experimental analysis; association between variables are often directed;
latent variables (e.g. eliteness) are integrated with manifest variables
(e.g. frequency); experimenters may investigate whether some variables
(e.g. frequency or eliteness) cause a change in other variables
(e.g. relevance). In the following, these properties are discussed.

\subsection{Variables and Covariation}
\label{sec:vari-covar}

\ac{IR} is naturally based on variables since the researchers can only come to
an understanding of how users and systems interact by using variables.  The
variables measured in \ac{IR} can be qualitative (e.g. class membership),
quantitative (e.g. term frequency), ordinal (e.g. document rank), cardinal
(e.g. document set size), integer or real.  Moreover, the variables are often
random, since some indexing and retrieval processes (e.g. relevance assessment)
are subject to uncertainty. In addition, covariation is the basis for many
retrieval and indexing models not only for finding term or document
correlations, but also for estimating the conditional probabilities that
are necessary for term weighting schemes such as \ac{BM25}.

The \ac{SEM}'s output provides evidence about whether the causal hypotheses of a
structural equation model such as $X \rightarrow Y$ can be confirmed by the data
collected from the manifest variables; for example, if the causal hypotheses of
a structural equation model are made between a variable measuring retrieval
effectiveness, $Y$, and variables describing the indexing and retrieval
processes dictated by a retrieval model, $X$, the \ac{SEM}'s output provides
evidence about whether $X$ can explain $Y$ and it may indicate some reasons that
a retrieval system performed badly (low $Y$) or satisfactorily (high $Y$) by
associating the values of $X$ to the values of $Y$.

Although covariation cannot be considered sufficient for convincing someone of a
causal relationship between two variables, it is nevertheless necessary in
\ac{IR} since it is unlikely that a causal relationship between two variables
(e.g. term frequency and pertinence) will occur without covariation.

The direction of the relationship between variables implies the distinction
between exogenous variables and endogenous variables.  Exogenous manifest
variables are usually frequencies, probabilities or sizes observed from the
collection indexes and aggregated at the level of topic or document; for
example, a variety of document statistics can be observed for each document and
then associated to its rank in a list of retrieved documents.  When evaluation
in \ac{IR} is considered, endogenous variables are usually referred to measures
of user satisfaction or document relevance; for example, retrieved document rank
is an endogenous manifest variable observed at the level of document and \ac{AP}
or \ac{NDCG} are endogenous manifest variables observed at the level of
topic. In this way, the variations of precision and recall can be explained by
the variations of exogenous or other endogenous variables.

\subsection{Endogenous Variables and Exogenous Variables}
\label{sec:endog-exog}

In general, endogenous variables are quite well distinguished from exogenous
variables in \ac{IR}.  The distinction between exogenous and endogenous
variables is made easier since it is possible to assign the role of exogenous
variables to documents and queries and the role of endogenous variables to
relevance assessments and retrieval effectiveness measures, for example.

Once endogenous variables and exogenous variables are assigned, an explanation
of the reasons why a system performs better than another system can be suggested
in terms of differences in the ways the exogenous variables are implemented by
the systems being compared. A richer description of the relationships between
variables can be obtained if additional factors explaining the reasons why the
exogenous variables may vary are added; for example, in interactive \ac{IR},
query expansion devices, relevance feedback algorithms and other methods
implementing user-document interaction may be considered as exogenous variables,
while measures other than precision such as user satisfaction or document
utility may integrate the endogenous variables; in Information Seeking, typical
endogenous variables have been the frequency of information sources used in
various groups \citep{Vakkari&05}.


\subsection{Latent Variables and Manifest Variables}
\label{sec:latent-manif}

While many variables, such as frequencies, are manifested because they can be
obtained by counting, other variables such as relevance and eliteness should be
viewed as latent. Relevance can be viewed as a latent variable because it
results from complex intellectual activities that cannot directly be
measured. However, relevance can indirectly be measured by means of manifest
variables that are considered signals or indicators of relevance. Relevance
labels or degrees are examples of relevance indicators because they can be
collected from human assessors or users, although they cannot represent the
context in which a document is deemed to be relevant.

Latent topics are another example of latent variables, since they are unobserved
terms, phrases or other textual sources that can indirectly be observed in the
form of (sequences of) words. \ac{LSA} which aims to discover latent topics in
the forms of word vectors by using unsupervised statistical methods
(e.g. \ac{SVD}) provides another example.

Data are usually raw in \ac{IR} since they are available as frequencies, scores,
and other numeric values. The problem with reproducing raw data is that
experimental systems often calculate weights and scores using different
parameters or methods, thus making the results slightly different.  If raw data
and exhaustive and precise documentation thereof were publicly available,
experiments might be reproduced.  Otherwise, covariance matrices are a compact
alternative to raw data. When covariance matrices are available, simulation or
meta-analysis can be performed, thus making experimental replication possible;
for example, a researcher may make his own covariance matrices available to the
research community, thus allowing the other researchers to reproduce experiments
and compare the experimental results without forcing them to reproduce the
experimental context and recollect the data.

The datasets used when \ac{SEM} is applied to \ac{IR} may store many records
since \ac{IR} experiments may produce large amounts of data from big test
collections. For example, the datasets used in this paper (see Section
\ref{sec:uses}) contains millions of documents, thousands of queries and
hundreds of features for each document-query pair.  When runs are utilised, it
is likely to be forced to process thousands of retrieved documents.  As \ac{SEM}
is a large sample methodology, its application to \ac{IR} does not pose a
problem.  Rather, some attention should be paid to the risk of easily rejecting
null hypotheses because of very large samples which may make any difference
significant.

The datasets mentioned above and especially the learning-to-rank datasets may
contain many features selected with the idea of providing the largest possible
amount of data to the researchers.  In that case, some variables may be highly
collinear. For example, term frequency and \acs{TFIDF} might be highly collinear
if the \acs{IDF} component discriminates terms very little.  We found a
significant number of highly collinear variables in the datasets used in the
experiments reported in Section \ref{sec:uses}.

In \ac{IR}, descriptive analysis is often performed by the researchers and
reported in the papers. For example, the variables that affect system
effectiveness measured by \ac{AP} are averaged and the variability across topics
can be described. Descriptive analysis tells what happens, however, it cannot
tell whether the variations observed are significant.  Another kind of analysis
can be the comparison between retrieval systems or components thereof; for
example, the \acp{MAP} of two competing retrieval systems can be compared and
the statistical significance of the difference between the \acp{MAP} can be
assessed in terms of p-value. A more complex analysis can be provided by success
and failure analysis, which provides evidence as to when a system fails to
retrieve relevant documents or succeeds in retrieving irrelevant documents
(reliability) or as to when the system failed to retrieve documents, tout-court
(retrievability).

\ac{SEM} is a complex of statistical procedures that may provide an explanation
of retrieval failures and successes because it allows the researcher to express
some hypotheses and test whether the observed data fit the model. If these
hypotheses were confirmed and if they were expressing reasons why a system
performed badly or worse than another system, we would be provided with a sound
methodology for diagnosis in \ac{IR} evaluation. It may be used to check whether
some general ideas underlying retrieval models are fitted by the data; for
example, this analysis will be used in Section \ref{sec:uses} to test whether
the observed data fit the structural equation model relating a combination of
authority and content to retrieval effectiveness, thus testing whether this
combination can help select relevant documents; another example will be the
structural equation model relating eliteness, term frequency and relevance and
underlying \ac{BM25}.

We mention two other types of structural equation models, i.e. \ac{CFA} models
and \ac{PA} models. The experiments that are reported in this paper are about
both types; however, the manifest variables are predominant in \ac{IR}. The
datasets of \ac{IR} experiments (e.g. test collections and learning-to-rank
datasets) usually include a number of manifest variables calculated from
documents, queries and user actions. In particular, the endogenous variables are
usually retrieval effectiveness values where the exogenous variables are
collection, document, query or user features. Many manifest variables are
interrelated and one variable may result from the other; for example, the sum of
\acsp{TF} in titles may determine the sum of \acsp{TF} in documents and
different \ac{PA} models may arise.

A beta coefficient is different from the correlation coefficient between two
variables. Suppose $X_1$ is term frequency, $X_2$ is click frequency, and $Y$ is
\ac{AP} and suppose $\cor{X_1}{Y} = 0.40$, $\cor{X_2}{Y} = 0.60$ and
$\cor{X_1}{X_2} = 0.60$. If the researcher excluded click frequency from the
structural equation model describing the relationships with \ac{AP}, he might
conclude that term frequency ($X_1$) positively determines retrieval
effectiveness and $\beta_1 = \cor{X_1}{Y}$. But if the researcher included click
frequency and investigated the structural equation model $\{X_1 \rightarrow Y,
X_2 \rightarrow Y\}$, the beta coefficient $\beta_1$ would reflect a different
relationship between term frequency and \ac{AP}, since $\beta_1 = (0.40 - 0.60
\cdot 0.60)/(1-0.60^2) = 0.06$, which is much lower than $\cor{X_1}{Y}$. The
reason is that the beta coefficient controls for the correlation between the
other predictors, whereas the correlation coefficient does not.

Moreover, the beta coefficients differ from the regression coefficients of a
structural equation model. Suppose that $Y$ is the \ac{AP} of a topic and $X$ is
the term frequency. When the covariance is positive, the value of $B$ would
indicate the predicted increase of performance measured by \ac{AP} for every
additional term occurrence. In contrast, standardized coefficients would
describe the effect of term frequency on performance in standard deviation
units, thus discarding the original scales of $X$ and $Y$. The beta coefficients
are instead necessary to compare the predictors within one structural equation
model, since they have the same standardized metric. For example, $\beta_2 =
(0.60 - 0.40 \cdot 0.60)/(1-0.60^2) = 0.56$, which is much greater than
$\beta_1$ relative to the difference between $\cor{X_1}{Y}$ and $\cor{X_2}{Y}$.


\subsection{Fitting Models and Data}
\label{sec:fitting-models-data}

When the variables that are relevant to a structural equation model are
specified and those manifest are collected and prepared, different kinds of
analysis can be performed. The simplest analysis is of descriptive nature, and
aims to select and summarize the data using statistical moments.  The kind of
analysis which we are interested to in this paper is confirmatory analysis:
given a structural equation model, confirmatory analysis tests whether the
observed data confirm (i.e. fit) the model.  Testing the structural equation
model would give evidence whether the causal hypotheses of the model can be
confirmed.

In \ac{SEM}, model comparison is implemented by chi-square difference statistic.
Chi-square is applied to hierarchical models, i.e. one is a proper subset of the
other.  For nonhierarchical models, \ac{AIC} or \ac{BIC} measure the information
loss when an experimenter would rather choose one model than choose another
model.

A structural equation model can be accepted as a valid model of the observed
data if these data fit the model.  The observed data fit a structural equation
model when there is no difference between the covariances predicted by the model
and the covariance estimated by the data.  When the difference is null, the fit
is exact. When the difference is almost null, the fit is close.  The null
hypothesis represents the researcher's hope that the structural equation model
fits the data, since the correspondence between the covariances predicted by the
model and the covariance estimated by the data means that the model describes
the data.  The rejection of the null hypothesis implies the rejection of the
structural equation model.

However, when the null hypothesis cannot be rejected, the null hypothesis cannot
be accepted. For example, sample size does matter; if one uses a very small
sample, any null hypothesis cannot be rejected, yet the null hypothesis -- the
structural equation model fits the data -- cannot be accepted.  This is the
reason that makes \ac{SEM} a ``large sample'' technique, since the failure of
rejecting the null hypothesis when the sample size is large should be a
``comfort evidence''.  It follows that the absence of statistical significance
(i.e. non-small p-value) supports he researcher's hope that the model is a good
fit, since the probability that more extreme fit indexes may be observed is not
small and the observed value is relatively close to the null hypothesis,
i.e. the data are consistent with the model.

\ac{RMSEA} is a fit index where a value of zero indicates the best result.
Actually, \ac{RMSEA} is reported as a confidence interval where the lower bound
is not negative.  When the hypothesis of exact fit is tested, the lower bound of
\ac{RMSEA} is greater than zero and the p-value is small (e.g.
$\mbox{p-value}<0.05$), it is unlikely that more extreme (i.e. smaller) values
of \ac{RMSEA} may be observed and the hypothesis of exact fit of the structural
equation model should be rejected. When the hypothesis of close fit is tested
and the p-value is not small (e.g. $\mbox{p-value}>0.10$), it is likely that
more extreme (i.e. larger) values of \ac{RMSEA} may be observed and the
hypothesis of close fit of the structural equation model should not be rejected.

Besides \ac{RMSEA}, approximate fit indexes such as \ac{CFI} \cite{Bentler90}
and \ac{TLI} \cite{Tucker73}, are continuous measures of correspondence between
the covariances predicted by the model and the covariance estimated by the data;
they may viewed as the degree to which the researcher's model is better than the
independence or baseline model.

\subsection{Further Explanations of the Differences between \acs{SEM} and other
  Techniques}
\label{sec:an-expl-diff}

In this section, we provide an example of comparison between techniques used to
analyzing experimental data.  Suppose a dataset stores one record for each
retrieved document with respect to a certain topic or query.  Such a record
contains a retrieval effectiveness measure (e.g. \ac{P@r}) and some features
(e.g. query-document term weights); the features are collected for each
retrieved document to allow researchers to analyze retrieval failures of one
topic or query at a time.  Alternatively, the dataset may contain one record for
each retrieval effectiveness measure at the level of topic or query; the
features are collected for each query to allow researchers to analyse the
overall retrieval effectiveness at the level of run.

The dataset should be processed to make variables as normal as possible and
eliminate outliers and collinearity.  Then, a correlation matrix can be
computed, for example:
\begin{equation}
  \nonumber
  \begin{array}{rrrrrrr}
    Y	&X_1	&X_2	&X_3	&X_4	&X_5	&X_6 \\
    1.0	&0.2	&0.3	&0.1	&-0.1	&0.2	&0.5\\
    0.2	&1.0	&0.9	&0.1	&0.0	&0.0	&0.0\\
    0.3	&0.9	&1.0	&0.0	&0.0	&0.0	&0.0\\
    0.1	&0.1	&0.0	&1.0	&0.0	&0.0	&0.0\\
    -0.1&0.0	&0.0	&0.0	&1.0	&0.4	&0.1\\
    0.2	&0.0	&0.0	&0.0	&0.4	&1.0	&0.9\\
    0.5	&0.0	&0.0	&0.0	&0.1	&0.9	&1.0
  \end{array}
\end{equation}
where $Y$ is a retrieval effectiveness measure and the $X_i$'s are the features.
Suppose the retrieval model utilised to generate the dataset promotes documents
when the query term weights increase.  The correlation matrix would help
experimenters to view the features that contribute more to retrieval
effectiveness than others.  However, $X_4$ is negatively correlated with $Y$,
thus suggesting a contrasting hypothesis, that is, the corresponding feature
makes retrieval effectiveness worse when the feature weight increases.

In point of fact, correlation coefficient hides the true relationship between
$Y$ and some features.  The change in standard deviations of $Y$, given a
1-point change in standard deviation of $X_4$, is about equal to $0.57$ (p-value
is close to zero).  The contrast between the $X_4$'s beta coefficient and
correlation coefficient is mainly due to the correlation between $X_4$ and $X_6$
and to that between $X_4$ and $X_7$.  When the beta coefficients are calculated
for each feature -- a regression model is estimated -- other contrasting results
are obtained.  The estimated regression model exhibits a good fit, since
$R^2=0.84$ and all the beta coefficients are highly significant.  In particular,
the $X_2$'s beta coefficient is about equal to $-0.36$ whereas the correlation
coefficient is $0.20$.  This contrast is mainly due to the correlation between
$X_2$ and $X_3$, which is very high.  Similarly, the $X_6$'s beta coefficient is
about equal to $-2.30$ because of the correlation with $X_7$.  In sum, beta
coefficients reveal the true impact of features on retrieval effectiveness.
Suppose the information about the $X$'s correlations is unavailable or
correlation is null.  Beta coefficients are equal to the corresponding
correlation coefficients -- and they can be of little help -- when the $X_i$'s
are uncorrelated.  Consider the following correlation matrix, for example:
\begin{equation}
  \nonumber
  \begin{array}{rrrrrrr}
    Y 	& X_1 	& X_2 	& X_3 	& X_4 	& X_5 	& X_6 \\
    1.0	& 0.2	& 0.3	& 0.1	& -0.1	& 0.2	& 0.5\\
    0.2	& 1.0	& 0.0	& 0.0	& 0.0	& 0.0	& 0.0\\
    0.3	& 0.0  	& 1.0	& 0.0	& 0.0	& 0.0	& 0.0\\
    0.1	& 0.0  	& 0.0	& 1.0	& 0.0	& 0.0	& 0.0\\
    -0.1	& 0.0  	& 0.0	& 0.0	& 1.0	& 0.0	& 0.0\\
    0.2	& 0.0  	& 0.0	& 0.0	& 0.0	& 1.0	& 0.0\\
    0.5	& 0.0  	& 0.0	& 0.0	& 0.0	& 0.0	& 1.0
  \end{array}
\end{equation}
Experimenters might be perplexed by the negative correlation between $Y$ and
$X_4$ if the feature was added to the model following the idea that an increase
of $X_4$ should cause and increase of $Y$.  Since the features are uncorrelated,
an explanation cannot be given in terms of the difference between beta
coefficients and correlation coefficients.

\ac{SEM} supports the experimenters and provides a solution.  To this end, the
experimenters have to define a structural equation model relating the features
to some latent variables, which may explain the negative correlation with
retrieval effectiveness.  For example, suppose an experimenter knows that $X_4$
and $X_6$ correspond to two query term weights of the same type (e.g. two query
term \acsp{IDF}) and s/he suspects that one term is about a query facet
complementary to the query facet of the other term.  A latent variable $A$ may
govern both features and cause the negative correlation.  The structural
equation model for this hypothesis can be written as follows:
\begin{equation}
  \nonumber
  Y \gets X_1+X_2+X_3+X_4+X_5+X_6 \qquad A \rightarrow X_4+X_6
\end{equation}
The fit of this model is good, since the p-value is about 0.63; therefore, the
model should not be rejected. \acf{CFI} is $1$; in other words, the approximate
fit is perfect; moreover, \acf{RMSEA} is zero and its confidence interval is
$[0,0.07]$, thus making the p-value of the hypothesis that \ac{RMSEA} is less
than $0.05$ equal to $0.84$.  \ac{SRMR} is also zero.  The beta coefficients
between $A$ and $X_4, X_6$ confirms the experimenter's hypothesis, since the
coefficients have opposite sign, in particular $A = 0.4X_4-0.2X_6$.  However,
the true nature of $A$ remains unknown, although it is certainly a numerical
feature.  It may refer to one term or to a set of terms -- discovering how
latent variables can be implemented is matter of future research.

As also explained in Section \ref{sec:sem-other-analysis}, \ac{SEM} differs from
other data analysis methods such as \acf{FA}.  \ac{FA} computes some factors
which are an alternative vector basis to the canonical vector basis underlying
the observed data.  The main advantage of \ac{FA} is the reduction of a large
set of variables to a small set of factors which approximate the correlation
matrix and then the relationships between variables.  Consider the correlation
matrix above.  The following factors can explain 73\% of variance:
\begin{equation}
  \nonumber
  \begin{array}{rrr}
    Z_1&Z_2&Z_3\\
    0.274& 0.204& 0.705 \\\
    & 0.896& \\
    & 1.002& \\
    &-0.183& 0.170 \\
    0.274& 0.134&-0.711 \\
    0.984&&-0.223 \\
    0.972&& 0.222 
  \end{array}
\end{equation}
The numbers are a measure of the contribution of a factor to a variable
and are called factor loadings.  \ac{FA} indicates that $Z_1$ influences
$X_5$ and $X_6$, $Z_2$ affects $X_1$ and $X_2$, and $Z_3$ influences $Y$
and $X_4$.  Clearly, the factors correspond to the main subsets of
related variables, yet their meaning is obscure, since $Y$ has been
considered in the same way as the $X$'s although the latter have been
considered exogenous variables in the structural equation model
above. However, the factors that are computed from a correlation matrix
cannot tell anything about the latent nature of unobserved variables.
Although the factor loadings may suggest that, say, $Z_3$ is a
``combination'' of $Y$ and $X_4$, which was found through an covariance
matrix approximation algorithm, it would be difficult to conclude that
it might be viewed as a meaningful variable.  Clearly, the researcher's
intervention would be necessary in the event that an interpretation were
useful.
\section{Using \acl{SEM} in \acl{IR} Evaluation}
\label{sec:uses}

In this section, we illustrate some applications of \ac{SEM} in \ac{IR}
evaluation.  In particular, we focussed on the comparison between retrieval
systems and on the latent variables that make retrieval effectiveness different;
for example, many retrieval systems fail in answering difficult queries -- those
for which precision is very low -- and experimenters need to know the causes of
failure.  However, what makes a query difficult might not make another query
difficult; therefore, two queries may require two different structural equation
models.  Although the structural equation models resulting from such analysis
are not the same, they can suggest some insights to the experimenter about how
the retrieval model should be modified in order to address the difficulty of the
queries.

Some datasets are needed for calculating the actual values of the manifest
variables.  The data used for this paper were derived by learning-to-rank
datasets and experimental retrieval results known as \emph{runs}; a run is a
data file storing the documents that are retrieved against each query and that
are ranked according to the degree of relevance.  In this paper, runs are joined
with \acp{qrel} to compute retrieval effectiveness measures.  Learning-to-rank
datasets describe documents and queries in terms of numerical features,
e.g. frequencies and lengths and \acp{qrel} at the level of document-query pair.

Only laboratory experiments based on experimental datasets were considered in
this paper.  Nevertheless, nothing in principle prevents from applying \ac{SEM}
to contexts other than laboratory, such as user studies or naturalistic studies
reported in Section \ref{sec:related}.

\subsection{Data Preparation}
\label{sec:data-preparation-1}

To be specific, we utilised two public learning-to-rank datasets:
\begin{itemize}
\item The \ac{LETOR} package (version 4.0) consists of three corpora and nine
  query sets as reported by \citeN{Qin&10}.  In our experiments, the Gov2 corpus
  and the 2007 Million Query track's query set were utilised.  Table
  \ref{tab:gov2-features} summarizes the 46 features of \ac{LETOR}.
\item The \ac{MSLR} package consists of two-large scale datasets. One dataset
  has 30,000 queries and 3,771,126 documents, the other dataset is a random
  sample. We utilised the random sample that has 10,000 and 1,200,193 documents.
  Table \ref{tab:mslr-vars} summarizes the features of \ac{MSLR} utilised in
  this paper.  \footnote{The complete list of features are available at
    \url{http://research.microsoft.com/en-us/projects/mslr/feature.aspx}.}
  \citeN{Liu11} reports further information.
\end{itemize}
The features of \ac{LETOR} and \acs{MSLR} were utilised to implement the
manifest variables of the structural equation models tested in the experiments
reported in this section.  However, before investigating some structural
equation models, the data were analysed as for collinearity and outliers.
\begin{table}
  \centering
  \scriptsize
  \begin{tabular}[t]{|r|l|l|}
    \hline
    Id	& Short name	& Feature description	\\
    \hline
    1	& bodytfsum	& $\sum_{t \in Q \cap D} \mbox{TF}(t,D)$ in body \\
    2	& anchortfsum	& $\sum_{t \in Q \cap D} \mbox{TF}(t,D)$ in anchor \\
    3	& titletfsum	& $\sum_{t \in Q \cap D} \mbox{TF}(t,D)$ in titlebody \\
    4	& urltfsum	& $\sum_{t \in Q \cap D} \mbox{TF}(t,D)$ in URL \\
    5	& tfsum		& $\sum_{t \in Q \cap D} \mbox{TF}(t,D)$ in $D$ \\
    6	& bodyidfsum	& $\sum_{t \in Q} \mbox{IDF}(t)$ in body \\
    7	& anchoridfsum	& $\sum_{t \in Q} \mbox{IDF}(t)$ in anchor \\
    8	& titleidfsum	& $\sum_{t \in Q} \mbox{IDF}(t)$ in titlebody \\
    9	& urlidfsum	& $\sum_{t \in Q} \mbox{IDF}(t)$ in URL \\
    10	& idfsum	& $\sum_{t \in Q} \mbox{IDF}(t)$ in $D$ \\
    11	& bodytfidfsum	& $\sum_{t \in Q \cap D} \mbox{TFIDF}(t,D)$ in body \\
    12	& anchortfidfsum& $\sum_{t \in Q \cap D} \mbox{TFIDF}(t,D)$ in anchor \\
    13	& titletfidfsum	& $\sum_{t \in Q \cap D} \mbox{TFIDF}(t,D)$ in titlebody \\
    14	& urltfidfsum	& $\sum_{t \in Q \cap D} \mbox{TFIDF}(t,D)$ in URL \\
    15	& tfidfsum	& $\sum_{t \in Q \cap D} \mbox{TFIDF}(t,D)$ in $D$ \\
    16	& bodydoclen	& $\sum_{t \in Q \cap D} \mbox{LENGTH}(D)$ in body \\
    17	& anchordoclen	& $\sum_{t \in Q \cap D} \mbox{LENGTH}(D)$ in anchor \\
    18	& titledoclen	& $\sum_{t \in Q \cap D} \mbox{LENGTH}(D)$ in titlebody \\
    19	& urldoclen	& $\sum_{t \in Q \cap D} \mbox{LENGTH}(D)$ in URL \\
    20	& doclen	& $\sum_{t \in Q \cap D} \mbox{LENGTH}(D)$ in $D$ \\
    21	& bodybm25	& $\sum_{t \in Q \cap D} \mbox{BM25}(t,D)$ in body \\
    22	& anchorbm25	& $\sum_{t \in Q \cap D} \mbox{BM25}(t,D)$ in anchor \\
    23	& titlebm25	& $\sum_{t \in Q \cap D} \mbox{BM25}(t,D)$ in titlebody \\
    24	& urlbm25	& $\sum_{t \in Q \cap D} \mbox{BM25}(t,D)$ in URL \\
    25	& bm25		& $\sum_{t \in Q \cap D} \mbox{BM25}(t,D)$ in $D$ \\
    26	& bodylmirabs	& $\sum_{t \in Q \cap D} \mbox{LMIRABS}(t,D)$ in body \\
    27	& anchorlmirabs	& $\sum_{t \in Q \cap D} \mbox{LMIRABS}(t,D)$ in anchor \\
    28	& titlelmirabs	& $\sum_{t \in Q \cap D} \mbox{LMIRABS}(t,D)$ in titlebody \\
    29	& urllmirabs	& $\sum_{t \in Q \cap D} \mbox{LMIRABS}(t,D)$ in URL \\
    30	& lmirabs	& $\sum_{t \in Q \cap D} \mbox{LMIRABS}(t,D)$ in $D$ \\
    31	& bodylmirdir	& $\sum_{t \in Q \cap D} \mbox{LMIRDIR}(t,D)$ in body \\
    32	& anchorlmirdir	& $\sum_{t \in Q \cap D} \mbox{LMIRDIR}(t,D)$ in anchor \\
    33	& titlelmirdir	& $\sum_{t \in Q \cap D} \mbox{LMIRDIR}(t,D)$ in titlebody \\
    34	& urllmirdir	& $\sum_{t \in Q \cap D} \mbox{LMIRDIR}(t,D)$ in URL \\
    35	& lmirdir	& $\sum_{t \in Q \cap D} \mbox{LMIRDIR}(t,D)$ in $D$ \\
    36	& bodylmirjm	& $\sum_{t \in Q \cap D} \mbox{LMIRJM}(t,D)$ in body \\
    37	& anchorlmirjm	& $\sum_{t \in Q \cap D} \mbox{LMIRJM}(t,D)$ in anchor \\
    38	& titlelmirjm	& $\sum_{t \in Q \cap D} \mbox{LMIRJM}(t,D)$ in titlebody \\
    39	& urllmirjm	& $\sum_{t \in Q \cap D} \mbox{LMIRJM}(t,D)$ in URL \\
    40	& lmirjm	& $\sum_{t \in Q \cap D} \mbox{LMIRJM}(t,D)$ in $D$ \\
    41	& pagerank	& PageRank of $D$ \\
    42	& inlinks	& Number of in-links of $D$ \\
    43	& outlinks	& Number of out-links of $D$ \\
    44	& urldepth	& Number of slashes of the $D$'s \ac{URL} \\
    45	& urllen	& Length of the $D$'s \ac{URL} \\
    46	& children	& Number of children of $D$ \\
    \hline
  \end{tabular}
  \caption{Features for the Gov2 corpus; for each feature, an
    identifier, a short name, and a description are
    provided. Symbols: $t$ is a term, $Q$ is a query, $D$ is a
    document. Notes: DIR = ``Dirichlet smoothing'', JM =
    ``Jelinek-Mercer smoothing'', ABS = ``Absolute discount
    smoothing''.} 
  \label{tab:gov2-features}
\end{table}

\begin{table}
  \centering
  \scriptsize
  \begin{tabular}[t]{|c|l|p{50mm}|}
    \hline
    Id & Short name & Description \\
    \hline
    1	& qtnbody & covered query term number body\\
    2	& qtnanchor & covered query term number anchor\\
    3	& qtntitle & covered query term number title\\
    4	& qtnurl & covered query term number url\\
    5	& qtn & covered query term number whole document\\
    12	& strmlenanchor & stream length anchor\\
    13	& strmlentitle & stream length title\\
    14	& strmlenurl & stream length url\\
    15	& strmlen & stream length whole document\\
    46	& tfnstrmlensumbody & sum of stream length normalized term frequency body\\
    47	& tfnstrmlensumanchor & sum of stream length normalized term frequency anchor\\
    48	& tfnstrmlensumtitle & sum of stream length normalized term frequency title\\
    50	& tfnstrmlensum & sum of stream length normalized term frequency whole document\\
    71	& tfidfsumanchor & sum of \ac{TFIDF} anchor\\
    73	& tfidfsumtitle & sum of \ac{TFIDF} title\\
    74	& tfidfsumurl & sum of \ac{TFIDF} url\\
    75	& tfidfsum & sum of \ac{TFIDF} whole document\\
    106	& bm25body & \ac{BM25} body\\
    107	& bm25anchor & \ac{BM25} anchor\\
    108	& bm25title & \ac{BM25} title\\
    109	& bm25url & \ac{BM25} url\\
    110	& bm25  & \ac{BM25} whole document\\
    111	& lmirabsbody & LMIR.ABS body (language model approach for \ac{IR} with absolute discounting smoothing)\\
    113	& lmirabstitle & LMIR.ABS title\\
    114	& lmirabsurl & LMIR.ABS url\\
    115	& lmirabs & LMIR.ABS whole document\\
    116	& lmirdiranchor & LMIR.DIR anchor (language model approach for \ac{IR} with Bayesian smoothing using Dirichlet priors)\\
    118	& lmirdirtitle & LMIR.DIR title\\
    119	& lmirdirurl & LMIR.DIR url\\
    120	& lmirdir & LMIR.DIR whole document\\
    126	& slashes & Number of slash in URL\\
    127	& urllen & Length of URL\\
    128	& inlink & Inlink number\\
    129	& outlink & Outlink number\\
    130	& pagerank & PageRank\\
    131	& siterank & SiteRank (Site level PageRank)\\
    132	& quality & QualityScore (the quality score of a web page; the score is outputted by a web page quality classifier)\\
    133	& badness & QualityScore2 (the quality score of a web page; the score is outputted by a web page quality classifier, which measures the badness of a web page)\\
    134	& query\_url\_clickcount & query-url click count (the click count of a query-url pair at a search engine in a period)\\
    135	& url\_clickcount & url click count (the click count of a url aggregated from user browsing data in a period)\\
    136	& url\_dwell\_time & url dwell time (the average dwell time of a url aggregated from user browsing data in a period)\\
    \hline
  \end{tabular}
  \caption{Exogenous manifest variables of the \ac{MSLR} dataset (query 22636) kept for the analysis. The complete list is available at {http://research.microsoft.com/en-us/projects/mslr/feature.aspx}.}
  \label{tab:mslr-vars}
\end{table}

If the analysis is performed at the level of query and not at the level of
document, the linked records can be grouped by run and query, and the features
are averaged for each group. Each resulting record was then linked to the
performance scores of the run for the query.
\begin{figure}[t]
  \begin{minipage}{1.0\linewidth}
    \includegraphics[width=\textwidth]{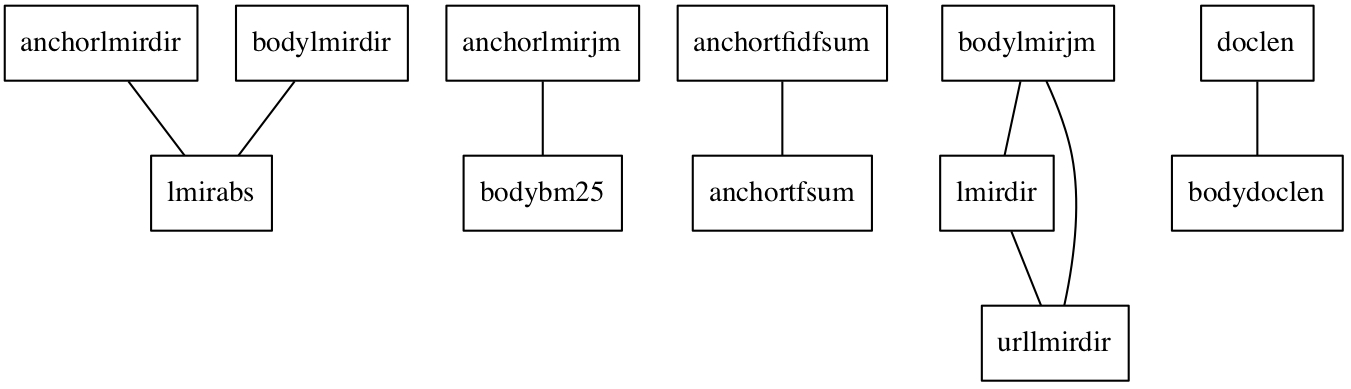}
    \\
    \includegraphics[width=\textwidth]{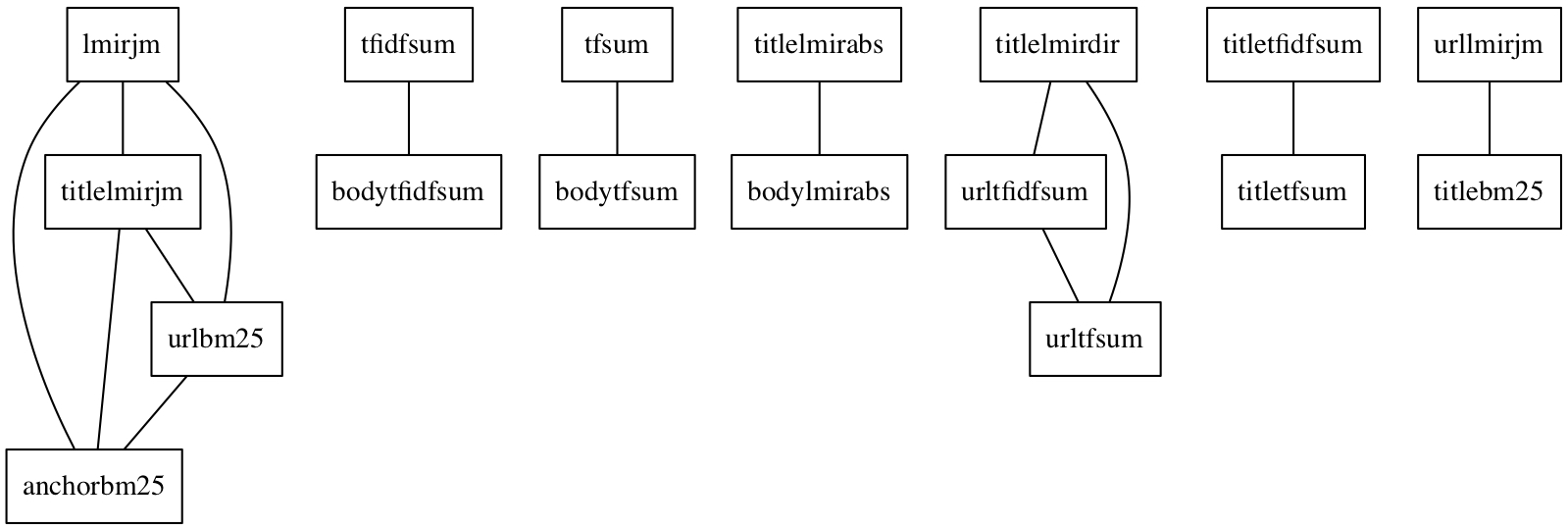}
  \end{minipage}
  \caption{Highly collinear variables in \ac{LETOR}. Each connected subgraph
  represents a subset of highly collinear variables. An edge was added when
  the Pearson correlation coefficient was 0.90 or more.}
  \label{fig:collinearity-1}
\end{figure}

We found high collinearity ($0.90+$) between some variables of both datasets.
A variable can thus be kept for each cluster of collinear variables and the
other variables can be ignored.  The removal of highly collinear variables might
be necessary, since the specific software tool used by a researcher to estimate
the parameters of a structural equation model may process complete covariance
matrices and not only the variables of the model.  We removed the most specific
features and kept the most general; for example, we kept term frequency within a
document and removed term frequency within the document title.
The criteria to ignore a variable depends on the ease of
interpretation of the \ac{SEM} results, since the results will not
significantly change with the ignored variables. Table
\ref{tab:collinearity} summarizes what was ignored and what was
kept.
\begin{table}
  \centering
  \scriptsize
  {\caption{Highly collinear variables kept for or ignored from analysis.
  \label{tab:collinearity}}{
  \begin{tabular}{|l|l|}
    \hline
    Variable kept & Variables ignored \\
    \hline
    anchortfsum & anchortfidfsum\\
    titlelmirdir & urltfidfsum, urltfsum\\
    titletfsum& titletfidfsum\\
    tfidfsum & bodytfidfsum\\
    tfsum & bodytfsum\\
    bodybm25 & anchorlmirjm\\
    lmirdir & bodylmirjm, urllmirdir\\
    lmirjm & titlelmirjm, urlbm25, anchorbm25\\
    bodylmirabs & titlelmirabs\\
    lmirabs & anchorlmirdir, bodylmirdir\\
    titlebm25&urllmirjm\\
    doclen & bodydoclen\\
    \hline
  \end{tabular}}
}
\end{table}

The outliers of a variable have been mapped to the mean value of the variable to
reduce the overall variability.  We also applied $\log (x + \min x + 1)$ to all
exogenous variables $x$ to reduce non-normality and variability of the
distribution of manifest exogenous variables and to make data closer to normal
distribution.  It is a standard practice in Statistics. There are other
transformations. Usually, a transformation improves how well a particular SEM
fits the data.

Kurtosis (i.e. heavier/lighter tails and a higher/lower peak than normal) and
skewness (i.e. asymmetry about normal mean) were reduced, yet not completely
eliminated.  However, ``children'' of \acs{LETOR} was still very skewed and
leptokurtic.  QQ-plotting allowed us to see that the lack of normality was due
to a very few large values while the others were null; this variable was then
ignored.  The other variables exhibit lack of normality at very high or very low
values. The middle values have a good fit with normality.

The manifest variables of \ac{LETOR} have well-scaled variances, since the scale
is 8:1, which is acceptable. Were the scale greater than hundreds, the values of
the variables exhibiting the smallest variance should be multiplied by a certain
factor until the scale becomes small.

Another approach can be based on reimplementing publicly documented
retrieval algorithms.  The difference between these two approaches lies in the
degree of control of the retrieval functions.  When using public datasets and
runs, the researcher investigates the retrieval functions designed and
implemented by other researchers, thus counting on the available documentation.
When reimplementing publicly documented retrieval algorithms, the researcher may
make decisions about some steps of indexing and retrieval which may make the
implemented retrieval functions slightly different from similar functions.  In
particular, the latter approach allows the researcher to investigate his own
retrieval functions.  The experiments that are reported in Section
\ref{sec:use-runs-letor} implemented the approach based on the reuse and
combination of public datasets.  In particular, two publicly available runs
submitted to the \ac{TREC} website and a public learning-to-rank dataset were
utilised.  We reproduced the runs obtained by a retrieval system based on
\ac{BM25} and those obtained by a retrieval system based on \ac{TFIDF} using the
TIPSTER test collection. In our experiments, the discs 4 and 5 of the TIPSTER
collection and the query sets of TREC-6, TREC-7 and TREC-8 were utilised to
perform the experiments.

\subsection{Use of Runs and \acl{LETOR} Datasets}
\label{sec:use-runs-letor}

Some structural equation models that are investigated in this paper include
endogenous variables based on precision.  Because precision is needed, document
ranking was necessary.  To obtain document ranking, we utilised two runs
submitted to the \ac{TREC} website for the 2007 Million Query track described by
\citeN{Allan&07}.  Two runs described by \citeN{Hiemstra&07} and produced using
a full-text index built by Lucene \citep{McCandless&10} were reused in our
experiments to generate the endogenous variables of some structural equation
models of this paper.  One run was based on the \ac{VSM} (\texttt{UAmsT07MTeVS})
and the other was based on the \ac{LM} (\texttt{UAmsT07MTeLM}).

\acl{LETOR} datasets were joined to the runs; in particular, for each run, every
query-document pair of a run was linked to the corresponding record of document
and query features.  Thus, we had one record of features for each run, query,
and document that can be in turn linked to the endogenous manifest variables
that can measure retrieval effectiveness; we utilised the ratio between the
numeric value of \ac{qrel} and document rank.  Each record of a run was joined
with the corresponding record of features; for example, each record of
\texttt{UAmsT07MTeVS} that refers to query $t$ and document $d$ was joined with
the record of \ac{LETOR} that refers to $t$ and $d$.  In this way, each record
of a run was an extended description of a retrieved document.  As for
\ac{LETOR}, we considered query 5440, for both runs, because the number of
retrieved documents was relatively high (about 80 documents), thus allowing us
to perform the experiments with a non-small sample.

\begin{figure}[t]
  \begin{minipage}{1.0\linewidth}
    \includegraphics[width=\textwidth]{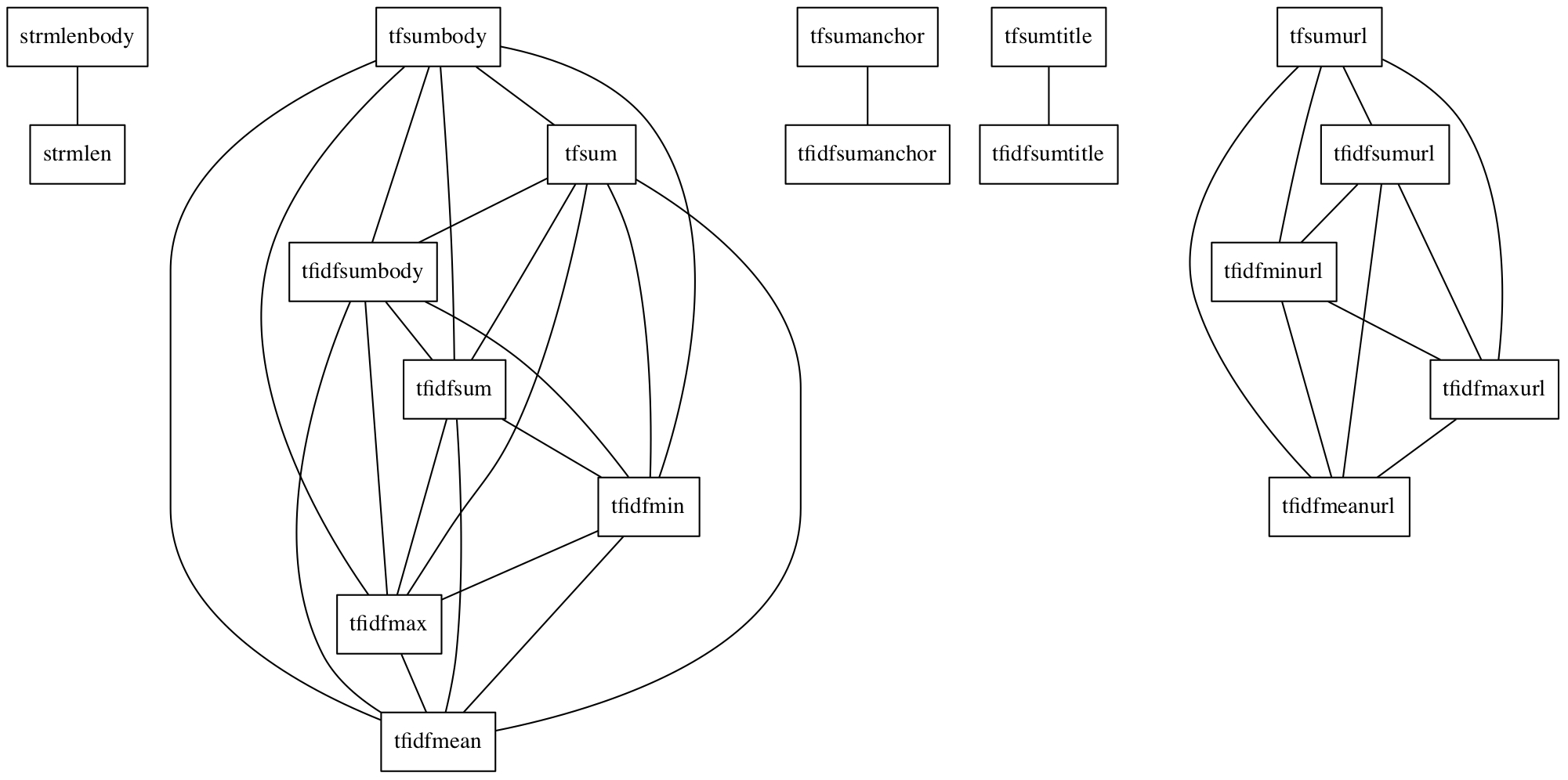}
    \\
    \includegraphics[width=\textwidth]{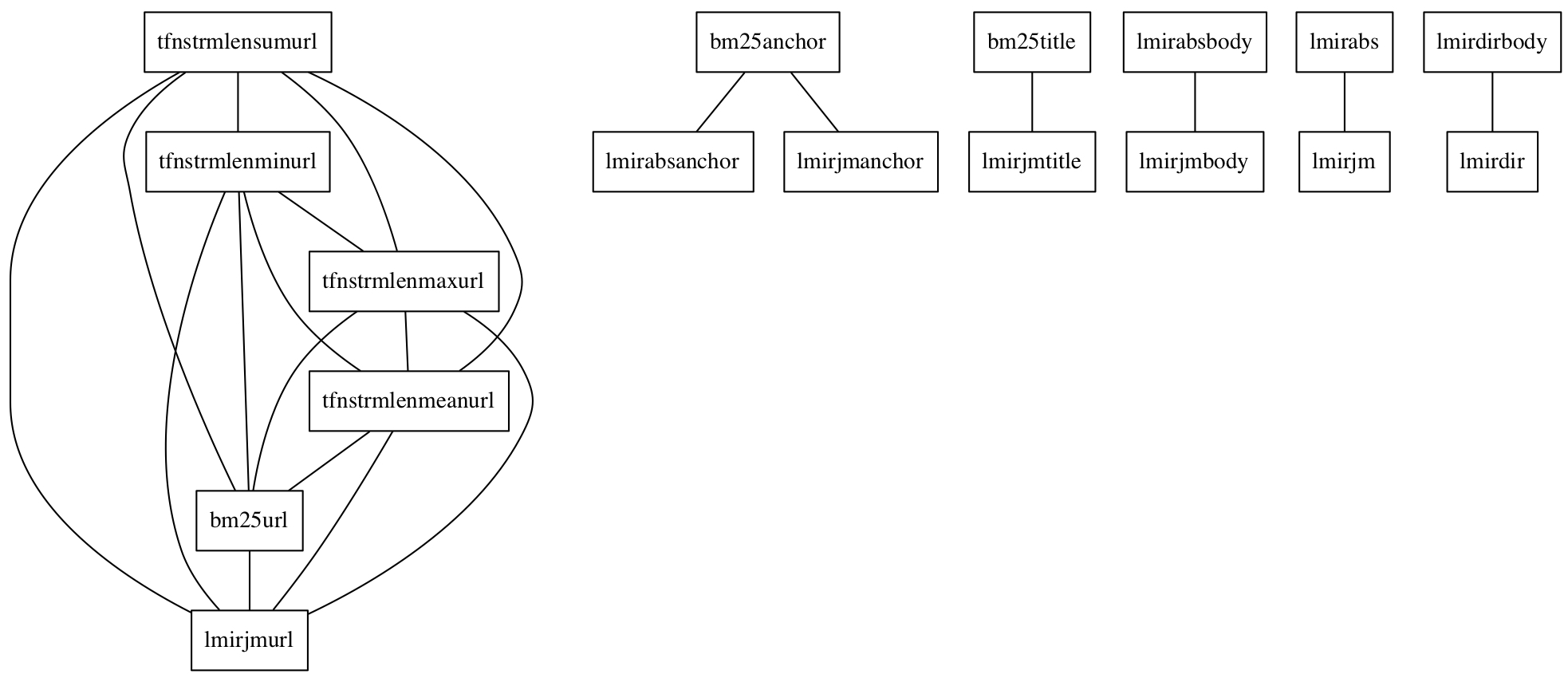}
  \end{minipage}
  \caption{Highly collinear variables in \ac{MSLR} for query 22636. Each
  connected subgraph represents a subset of highly collinear variables. An
  edge was added when the Pearson correlation coefficient was 0.975 or
  more.}
  \label{fig:collinearity-2}
\end{figure}

The \ac{MSLR} dataset was investigated through query 22636, which is related to
a relatively high number of cases (809) and all the five relevance degrees were
assigned to the cases.  Two variables were highly collinear if the correlation
was 0.975 or more \citep{Kline15}. Some methods are suggested in the literature,
yet thresholds are empirically chosen.  Something similar happens with p-values
which are compared with standard threshold (e.g. 0.01 or 0.05).  In the paper,
the thresholds were chosen by visual inspection; the threshold was the minimum
value that induces disconnected and complete subgraphs of manifest variables as
depicted by Figs. \ref{fig:collinearity-1} and \ref{fig:collinearity-2}.  The
clusters of highly collinear variables are depicted in
Fig. \ref{fig:collinearity-2}.  After ignoring the highly collinear variables,
the variables involved during the analysis are reported in Table
\ref{tab:mslr-vars}.  To reduce lack of normality of the remaining variables,
the transformation $\log(x+min(x)+1)$ was applied to all exogenous variables.
Unlike \ac{LETOR}, the ratio between the maximum variance and the minimum
variance was very high (12/0.001) in \ac{MSLR}.  As the large distance between
maximum variance and minimum variance would cause problems during parameter
estimation, it was progressively reduced by doubling the variable with minimum
variance until the ratio was not greater than 10.  (see the algorithm of
Fig. \ref{fig:scalevar}).

\begin{table*}[t]
  \centering
  \scriptsize
  \begin{tabular}[t]{lrrrrrrrrrrrrrrr}
    \hline
    {}&1&2&3&4&5&6&7&8&9&10&11&12&13&14&15\\
    \hline
    qtnbody&-1&0&0&0&0&0&1&0&0&0&0&0&0&1&0\\
    qtnanchor&0&-1&-1&1&0&0&0&0&0&0&0&1&0&0&-1\\
    qtntitle&-1&0&0&0&0&0&0&1&0&-1&0&0&-1&1&0\\
    qtnurl&0&-1&1&0&-1&0&0&0&0&-1&0&0&-1&-1&1\\
    qtn&-1&0&0&0&0&0&1&0&0&0&-1&-1&-1&0&0\\
    strmlenanchor&0&-1&-1&0&0&0&0&0&0&0&-1&-1&0&0&0\\
    strmlentitle&-1&0&0&-1&0&-1&-1&-1&-1&1&-1&0&0&0&1\\
    strmlenurl&0&-1&0&-1&1&1&0&1&1&1&-1&0&-1&0&0\\
    strmlen&-1&0&0&-1&0&-1&0&0&0&0&0&0&0&0&0\\
    tfnstrmlensumbody&-1&0&0&1&0&1&0&0&1&1&1&0&1&0&1\\
    tfnstrmlensumanchor&0&-1&-1&1&0&0&0&0&0&0&0&1&0&0&0\\
    tfnstrmlensumtitle&-1&0&0&0&1&1&-1&1&1&-1&1&0&0&0&0\\
    tfnstrmlensum&-1&0&1&1&0&1&0&-1&0&1&0&0&1&-1&0\\
    tfidfsumanchor&0&-1&-1&1&0&0&0&0&0&0&0&0&0&0&0\\
    tfidfsumtitle&-1&0&0&0&0&0&-1&0&0&0&0&0&0&0&0\\
    tfidfsumurl&0&-1&1&0&-1&0&0&0&0&-1&0&0&0&0&0\\
    tfidfsum&-1&0&0&0&0&-1&0&-1&0&1&0&0&0&0&0\\
    bm25body&-1&0&0&0&0&0&1&0&0&0&0&0&0&1&0\\
    bm25anchor&0&-1&-1&1&0&0&0&0&0&0&0&1&0&0&0\\
    bm25title&-1&0&0&0&0&0&-1&1&0&-1&0&0&0&0&0\\
    bm25url&0&-1&1&0&-1&0&0&0&0&-1&0&0&0&0&1\\
    bm25&-1&0&0&0&0&0&1&0&0&0&-1&0&0&0&0\\
    lmirabsbody&-1&0&0&0&0&0&1&0&0&0&0&0&0&0&0\\
    lmirabstitle&-1&0&0&0&1&0&-1&1&0&-1&0&0&0&0&0\\
    lmirabsurl&0&-1&1&0&-1&0&0&0&-1&-1&1&0&0&0&0\\
    lmirabs&-1&0&1&1&0&1&1&0&0&0&-1&0&0&0&0\\
    lmirdiranchor&0&-1&-1&0&-1&0&0&1&0&0&1&-1&1&1&1\\
    lmirdirtitle&-1&0&0&0&0&0&-1&-1&-1&1&-1&0&1&-1&0\\
    lmirdirurl&0&-1&1&-1&-1&0&0&1&-1&1&0&-1&1&1&-1\\
    lmirdir&-1&0&0&0&0&-1&0&-1&0&1&0&0&0&0&0\\
    slashes&0&-1&0&-1&1&1&0&0&1&0&-1&0&0&0&0\\
    urllen&0&-1&0&-1&1&0&0&1&1&1&-1&0&-1&1&0\\
    inlink&0&-1&-1&0&1&1&0&-1&0&-1&1&-1&0&-1&-1\\
    outlink&0&-1&0&-1&0&-1&0&-1&1&1&1&0&0&-1&0\\
    pagerank&0&0&-1&-1&-1&1&-1&-1&-1&0&1&0&-1&1&0\\
    siterank&0&-1&-1&-1&0&1&0&-1&0&1&1&1&-1&0&0\\
    quality&1&0&0&1&0&0&-1&0&0&1&1&0&-1&1&0\\
    badness&0&0&1&1&0&-1&-1&-1&0&1&0&-1&-1&0&-1\\
    query\_url\_clickcount&0&0&0&0&-1&0&-1&-1&1&-1&-1&0&1&1&0\\
    url\_clickcount&0&1&-1&-1&-1&0&0&1&1&1&-1&-1&-1&-1&0\\
    url\_dwell\_time&0&1&-1&0&-1&0&0&1&0&0&-1&-1&-1&-1&0\\
    \hline
  \end{tabular}
  \caption{The first 16 principal components of the \ac{MSLR} rescaled variables for query 22636}
  \label{tab:mslr-t22636-princomp}
\end{table*}

\begin{figure}[t]
  \centering
  \begin{algorithmic}
    \REQUIRE Dataset of $k$ manifest variables, $X_1,\dots,X_k$
    \STATE $\text{illscaled} \gets \text{TRUE}$
    \WHILE{$\text{illscaled}$}{
      \STATE $i_{\max} \gets \arg_{i=1,\dots,k}\max{\var{X_i}}$
      \STATE $i_{\min} \gets \arg_{i=1,\dots,k}\min{\var{X_i}}$
      \IF{$\var{X_{i_{\max}}} / \var{X_{i_{\min}}} \leq 10$}{
        \STATE $\text{illscaled} \gets \text{FALSE}$
      }
      \ELSE{
        \STATE $X_{i_{\min}} \gets 2 X_{i_{\min}}$
      }
      \ENDIF
    }
    \ENDWHILE
  \end{algorithmic}
  \caption{The algorithm used to rescale the variables until the variances were no longer ill-scaled.}
  \label{fig:scalevar}
\end{figure}
In the following sections, some analysis of experimental retrieval
results have been illustrated.

\subsection{Testing What Affect Effectiveness}
\label{sec:testing-effect}

Consider the manifest variables of \texttt{UAmsT07MTeVS} and
\texttt{UAmsT07MTeLM} after applying the logarithmic transformation to reduce
non-normality.

As mentioned above, the endogenous variable was the ratio between the numeric
value of \ac{qrel} and document rank.  In order to reduce the variability, a
log-transformation was applied to this ratio too.  As the numeric value of
\ac{qrel} may be zero and a log-transformation cannot be applied to zero, the
actual transformation was $Y = \log(\mbox{qrel}+1)/(\mbox{rank}+1)$ where qrel
is the numeric value of \ac{qrel}.  The argument of the logarithmic function is
positive when the document at the rank of the denominator is relevant and
decreases when rank increases. It is a precision measure at the level of
document since it is the contribution of a document to precision. Instead,
\ac{P@r} is a measure of precision at the level of document list since it is the
precision while the sublist of the top $r$ documents is scanned. The $Y$ defined
above is preferable to \ac{P@r} because the analysis performed on
\texttt{UAmsT07MTeVS} and \texttt{UAmsT07MTeLM} was at the level of document --
the \ac{LETOR} records were indeed joined to documents and not to lists.

After preparing the data, we looked for the best path model fitting the
exogenous variables to the endogenous variable that measure retrieval
effectiveness.  To this end, a process of experimenting with various path models
was performed until a good fit was found.  In the experiments of this paper, the
path model for \texttt{UAmsT07MTeVS} was $Y \gets \log (\mbox{bodybm25}+1) +
\log (\mbox{titlebm25}+1) + \log (\mbox{anchortfsum}+1)$ and that for
\texttt{UAmsT07MTeLM} was $Y \gets \log (\mbox{bodybm25}+1) + \log
(\mbox{lmirjm}+1) + \log (\mbox{tfsum}+1)$.  Two structural equation models have
been tested in the experiments:\footnote{The intercepts were removed because
  they were of little significance.}
\begin{eqnarray*}
  Y &=& B_{\mbox{\scriptsize bodybm25}} \log (\mbox{bodybm25}+1) +\\
  && B_{\mbox{\scriptsize titlebm25}} \log (\mbox{titlebm25}+1) +\\
  && B_{\mbox{\scriptsize anchortfsum}} \log (\mbox{anchortfsum}+1)
\end{eqnarray*}
for \texttt{UAmsT07MTeVS} and
\begin{eqnarray*}
  Y &=& B_{\mbox{\scriptsize bodybm25}} \log (\mbox{bodybm25}+1) +\\
  &&    B_{\mbox{\scriptsize lmirjm}} \log (\mbox{lmirjm}+1) +\\
  && B_{\mbox{\scriptsize
      tfsum}} \log (\mbox{tfsum}+1)
\end{eqnarray*}
for \texttt{UAmsT07MTeLM}.  An exogenous variable was significant
when its regression coefficient was statistically significant
($\mbox{p-value} \approx 0$); the variables of the two structural
equation models have significant regression coefficients, in
particular,
\begin{eqnarray*}
  B_{\mbox{\scriptsize bodybm25}} = 6.55
  \\ 
  B_{\mbox{\scriptsize titlebm25}} = 5.10
  \\ 
  B_{\mbox{\scriptsize anchortfsum}} = -1.67
\end{eqnarray*}
for \texttt{UAmsT07MTeVS} and
\begin{eqnarray*}
  B_{\mbox{\scriptsize tfsum}} = 1.89 
  \\ 
  B_{\mbox{\scriptsize lmirjm}} = 9.10
  \\
  B_{\mbox{\scriptsize bodybm25}} = 1.45
\end{eqnarray*}
for \texttt{UAmsT07MTeLM}.  The beta coefficients are
\begin{eqnarray*}
  \beta_{\mbox{\scriptsize bodybm25}} = 0.44
  \\ 
  \beta_{\mbox{\scriptsize titlebm25}} = 0.64
  \\ 
  \beta_{\mbox{\scriptsize anchortfsum}} = -0.10  
\end{eqnarray*}
for  \texttt{UAmsT07MTeVS} and
\begin{eqnarray*}
  \beta_{\mbox{\scriptsize tfsum}} = 0.15
  \\ 
  \beta_{\mbox{\scriptsize lmirjm}} = 0.87 
  \\
  \beta_{\mbox{\scriptsize bodybm25}} = 0.10
\end{eqnarray*}
for \texttt{UAmsT07MTeLM}, thus confirming the role played by \ac{BM25} -- early
introduced by \cite{Robertson&94} -- and \ac{LM} -- proposed by \cite{Ponte&98}
-- for these two runs.
Using the proportion of variance explained by all manifest variables with direct
effects on the endogenous variable, we have a measure of goodness-of-fit
$R^2$. The $R^2$'s values of the two models were $0.85$ and $0.90$ respectively
for \texttt{UAmsT07MTeLM} and \texttt{UAmsT07MTeVS}, thus suggesting a good fit
of the endogenous variable.

Finding the best path models was not a straightforward process.  Indeed, given
the endogenous variable, a path model is defined on the basis of a subset of
exogenous variables, therefore, the best path model was the subset of exogenous
variables that best fit the endogenous variable.
Moreover, the process to find the best fit is manual and based on the
researcher's knowledge of the application domain.  The difficulty of finding the
best fit is hampered by the potential complete enumeration all the possible
subsets, whose exponential number is $2$ to the power of the number of exogenous
variables, the latter requiring an infeasible amount of work even for not large
numbers.  To cope with this exponential order, the semantics of the exogenous
variables and the description of the retrieval algorithm utilised to produce a
run helped select the most appropriate variables; for example, pagerank, which
was computed by the PageRank algorithm introduced by \cite{Brin&98}, is unlikely
to correlate with effectiveness when \texttt{UAmsT07MTeVS} is considered,
whereas tfsum would be more appropriate.  Although the researcher's knowledge of
the application domain seems necessary to limit the space of subsets of
exogenous variables, it is still likely that some subsets might be missed, thus
making the selected structural equation models less than optimal.

As for \texttt{UAmsT07MTeLM}, the type of smoothing plays a crucial role because
the effectiveness of the exogenous variable explaining the endogenous variable
changes with smoothing technique.  Indeed, $R^2$ significantly decreases if
lmirjm is replaced with lmirdir or lmirabs, the latter being an outcome
explained by the negative correlation between lmirjm and lmirdir
($\mbox{p-value} < 0.05$) and that between lmirjm and lmirabs ($\mbox{p-value} <
0.01$).

In contrast, the importance of bodybm25, which is the backbone of the
probabilistic models, for the \ac{VSM}-based run is worth noting especially if
it is compared with the importance of the variables significantly related to the
\ac{VSM} such as anchortfsum.
However, the important role played by bm25 should not come as a complete
surprise.  The sum of \acs{TFIDF} weights (tfidfsum) provided by \ac{LETOR} has
been computed using a mathematical formulation different from the formulation
implemented by modern \ac{VSM} retrieval systems such as Lucene, which was used
in the experiments reported by \citeN{Hiemstra&07}.  Indeed, the Lucene
formulation is more similar to \ac{BM25} than to the \ac{LETOR}'s tfidfsum, thus
explaining why bodybm25 explains retrieval effectiveness in the \ac{VSM}-based
run.  The small statistical correlation between bodytfidfsum and bodybm25 has
further confirmed that their mathematical formulations were different.  The main
reason for this discrepancy was due to doclen, which is the most correlated
variable with both bodytfidfsum and bodybm25 (both p-values were not greater
than 0.01): the correlation between doclen and bodytfidfsum was positive,
whereas that between doclen and bodybm25 was negative.

The role played by \ac{BM25} in the \ac{VSM}-based run mentioned above might be
considered an example of what \ac{SEM} can suggest when applied to investigate
experimental results.  Some variables that are absent from a ranking function
may have a role in a revised ranking function because they are significantly
related to retrieval effectiveness in so far as its beta coefficient suggests.
The revised ranking function may include the new variable using some
mathematical or algorithmic rule decided by the researcher hoping that the new
variable can boost the ranks of retrieved relevant documents or the retrieval of
additional relevant documents.

The goodness-of-fit changes when the \ac{LM} scores utilised as exogenous
variables are those calculated from document parts other than the complete
document; for example, if lmirjm is replaced with bodylmirjm, $R^2$
decreases. Similarly, the effectiveness of \ac{BM25} in explaining the
endogenous variable of \texttt{UAmsT07MTeVS} depends on the document part from
which the estimation data are extracted; for example, when bodybm25 is replaced
with bm25 the goodness-of-fit decreases considerably, thus suggesting that the
distribution of the terms significantly changes when it is estimated from
different document parts.

Structural equation modeling depends on query and on run; indeed, testing the
models found for query 5440 and applied to \texttt{UAmsT07MTeLM} and
\texttt{UAmsT07MTeVS} for another query (e.g. 2297) gave unsatisfactory results
as shown by the significant decrement of $R^2$.  This outcome and the
dependencies of the document parts from which estimation is performed are both
an issue and a strength of the \ac{SEM}-based approach to diagnose \ac{IR}
evaluation.  On the one hand, it is an issue because a structural equation model
has to be found for each retrieval algorithm (i.e. run) and for each query, and
finding such a model requires an intellectual effort of the researcher who has
to apply his expertise in the application domain being investigated by means of
\ac{SEM} for each run and query.  On the other hand, the adaptation of the
structural equation model to both run and query can provide an in-depth
description of the retrieval system's performance for each query and can make
the failure analysis at the level of query possible and effective.  Such a
dependency calls from fully or semi-automatic methods for generating and testing
structural equation models that can support the \ac{IR} researcher in analysing
the successes and the failures of a retrieval system.

\subsection{Testing Latent Variables Behind Manifest Variables}
\label{sec:test-latent-perf}

In \ac{IR}, researchers often assume the presence of latent variables such as
relevance, authoritativeness (introduced by \cite{Brin&98} and
\cite{Kleinberg99}) and eliteness (introduced by \cite{Harter75a}) behind the
observed variables such as term frequencies and \acp{qrel}. For example, the
fact that relevance cannot be reduced to aboutness and that further dimensions
of relevance such as document authoritativeness and quality should be considered
in a retrieval function is by now well accepted. Another example is the
metaphor of the \ac{LM} approach introduced by \citeN{Ponte&98}. It assumes that
both the authors of a document and the users who assess the document as relevant
write the document and queries, respectively, that are about the same query,
thus establishing a relationship between relevance and aboutness.

Another case of structural equation model including latent variables describes
the relationship between eliteness, relevance and term frequency hypothesized by
\citeN{Robertson&09a} in the context of \ac{BM25} and stemmed from the intuition
given by \citeANP{Harter75a} \citeyear{Harter75a,Harter75b} that term eliteness
can be related to relevance.  According to the relationship between eliteness,
relevance and term frequency, for any document-term pair, eliteness is a latent
property such that if the term is elite, then the document is about the concept
represented by the term.  Eliteness cannot be observed directly because it a
latent variable.  Manifest variables such as \acp{qrel} and term frequency are
indirect manifestations of eliteness.  Specifically, this relationship can be
described as follows: (1) eliteness is a property of a document-term pair such
that if the term is elite in the document, in a sense the document is about the
concept denoted by the term; (2) aboutness is a property of a document-query
pair such that if the document is about a concept denoted by a query term, the
document is relevant to the information need described by the query. These
relations would be enough to explain the association between term frequency and
relevance to the query.  The relationship between eliteness, aboutness and
relevance can be viewed as an example of what has been illustrated in this
paper, that is, how relationships of this kind can be formalised as linear
equations relating exogenous or endogenous variables and manifest or latent
variables using a methodology based on a relatively simple mathematical model.

Using \ac{SEM}, it is possible to formulate the eliteness model
using the following structural equation model:
\begin{eqnarray}
  \label{eq:eliteness}
  \mbox{qrel} \rightarrow \mbox{eliteness} \rightarrow \log \mbox{tfsum}
\end{eqnarray}
where eliteness is a latent variable whereas tfsum and qrel are manifest
variables.  This model postulates that relevance ``causes'' eliteness that in
turn ``causes'' tfsum.  Using \texttt{UAmsT07MTeVS} for query 5440, the
relationship between qrel and eliteness is not significant
($\mbox{p-value}=0.637$), whereas that between tfsum and eliteness is
significant ($\mbox{p-value} \approx 0$).  This structural equation model and
the regression coefficients thereof should not be rejected according to the
chi-square test ($\mbox{p-value} = 0.637$).  The empirical numbers for CFI and
RMSEA are 1 and 0, respectively, because  the number of observations equals
the number of parameters.  In the example, the parameters are the direct effects
(arrows) on endogenous variables (eliteness, log(tfsum)) as well as the
variances of the exogenous variable (qrel). The observations of the model are
the variances and covariances of the manifest variables (qrel, log(tfsum)).

The suggestion that the structural equation model \eqref{eq:eliteness} should
not be rejected means that the hypothesis that eliteness is not significantly
caused by relevance should not be rejected, thus not confirming the hypothesis
made by \citeN{Robertson&09a}.  However, the following slightly different
structural equation model
\begin{eqnarray}
  \label{eq:eliteness-2}
  \mbox{qrel} \rightarrow \mbox{eliteness} \rightarrow \log \mbox{bm25}
\end{eqnarray}
suggests that that hypothesis should be considered because the
relationship between eliteness and qrel is significant.

Tests of the structural equation model \eqref{eq:eliteness} were replicated over
all the queries in the \ac{LETOR} data set to show that \ac{SEM} can be applied
to the level of runs as well.  The documents retrieved by \texttt{UAmsT07MTeLM}
were first joined with their features and then grouped by query.  For each
query, the average value of each feature and the \ac{NDCG} value for that query
were computed, thus obtaining a data set at the level of query.  The fit of the
structural equation model was extremely good.  We found that tfsum can be
determined by eliteness because the regression coefficient was about 0.35 and
the p-value was about zero, but qrel does not determine eliteness because the
regression coefficient was not significant. The p-value of the model was about
0.6, thus it cannot be rejected.
Similar results were obtained by testing the structural equation model
\eqref{eq:eliteness-2} or by replacing qrel with \ac{NDCG}, which was introduced
by \cite{Jarvelin&02}.

Another example of the use of \ac{CFA} in \ac{IR} is the investigation of the
coexistence of authoritativeness and aboutness as two distinct latent variables
in the same document. A document can be viewed as authoritative when is able to
be trusted as being accurate, true or reliable; other terms that are used to
describe this document features are credibility or veracity; in contextual
\ac{IR}, authoritativeness can be viewed as a factor of document quality
\cite{Melucci12b} and can be measured by, for example, PageRank.  The following
model in which \ac{qrel} can be a manifestation of both latent variables may
model the coexistence of authoritativeness and aboutness as two distinct latent
variables in the same document:
\begin{eqnarray}
  \mbox{authoritativeness} &\rightarrow& \mbox{pagerank} +
  \mbox{indegree} + \\ && \mbox{urldepth} + \mbox{qrel}
  \label{eq:auth-about-1} \\
  \mbox{aboutness} &\rightarrow& \mbox{doclen} + \mbox{bm25} + \mbox{qrel}
  \label{eq:auth-about-2} \\
  \mbox{authoritativeness} &\leftrightarrow& \mbox{aboutness}
  \label{eq:auth-about-3}
\end{eqnarray}
where the regression coefficients of \eqref{eq:auth-about-1} are
\begin{eqnarray*}
  B_{\mbox{\scriptsize pagerank}} &=& 0.044 \\ B_{\mbox{\scriptsize
      indegree}} &=& 0.006 \\ B_{\mbox{\scriptsize urldepth}} &=& -0.043
  \\ B_{\mbox{\scriptsize qrel}} &=& 0.022
\end{eqnarray*}
and the regression coefficients of \eqref{eq:auth-about-2} are
\begin{eqnarray*}
  B_{\mbox{\scriptsize doclen}} &=& 0.137 \\  B_{\mbox{\scriptsize
      bm25}} &=& -0.047 \\  B_{\mbox{\scriptsize qrel}} &=& 0.396
\end{eqnarray*}
These coefficients are statistically significant with $\mbox{p-value}
< 0.01$ except for $B_{\mbox{\scriptsize qrel}}$ of
\eqref{eq:auth-about-1}.  The correlation between authoritativeness
and aboutness is insignificant.  This model passes the chi-square
exact-fit test ($\mbox{p-value} = 0.116$) as confirmed by
$\mbox{\ac{CFI}} = 0.934$.  It also passes the approximate fit test
since $\mbox{\ac{RMSEA}} = 0.089$ ($\mbox{p-value} = 0.212$).

Latent variables were investigated also using \ac{MSLR}.  Besides including many
more variables than \ac{LETOR}, \ac{MSLR} also includes variables about the
behaviour of the users who visited the pages described in the dataset and a
couple of variables about the quality of the pages visited by the users.  The
variety of manifest variables of \ac{MSLR} allowed us to make some hypotheses
about the latent variables that may affect retrieval effectiveness.  In
particular, it was hypothesized that four latent variables, i.e. content, link,
graph, page and user, may explain the manifest endogenous variable named
``qrel'' that encodes retrieval effectiveness (qrel ranges from 0 to 4).  The
latent variable ``content'' was about the informative content (i.e. keywords) of
the pages that matched the query's informative content.  The latent variable
``link'' was about the informative content stored in the \acp{URL} and in the
link anchors that matched the query's informative content.  The latent variable
``graph'' was about the graphical properties of the \ac{WWW} node that
corresponds to the page.  The latent variable ``user'' was about the behaviour
of the user who visited the page. The latent variable ``page'' was about the
quality of the page.  Thus, we have the following structural equation model:
\begin{eqnarray*}
  \mbox{qrel} &\gets& \mbox{content}+\mbox{link}+\mbox{graph}+\mbox{quality}+\mbox{user} \\
  \mbox{content} &\rightarrow& \mbox{qtnbody} + \mbox{qtntitle} + \mbox{qtn} + \mbox{strmlentitle} + \\
  && \mbox{strmlen} + \mbox{tfnstrmlensumbody} + \mbox{tfnstrmlensumtitle} \\
  && + \mbox{tfnstrmlensum} + \mbox{tfidfsumtitle} + \mbox{tfidfsum} + \\
  && \mbox{bm25title} + \mbox{bm25} + \mbox{lmirabsbody} + \\
  && \mbox{lmirabstitle} + \mbox{lmirabs} + \mbox{lmirdirtitle}\\
  \mbox{graph} &\rightarrow& \mbox{pagerank} + \mbox{inlink} + \mbox{outlink}
  + \mbox{siterank} \\
  \mbox{link} &\rightarrow& \mbox{qtnurl} + \mbox{strmlenanchor} + \mbox{strmlenurl} \\
  && + \mbox{tfnstrmlensumanchor} + \mbox{tfidfsumanchor} + \mbox{tfidfsumurl}\\
  && + \mbox{bm25anchor} + \mbox{bm25url} + \mbox{lmirabsurl} + \\
  && \mbox{lmirdiranchor} + \mbox{lmirdirurl}\\
  \mbox{page} &\rightarrow& \mbox{quality}+\mbox{badness} \\
  \mbox{user} &\rightarrow& \mbox{query\_url\_clickcount}+\mbox{url\_clickcount}
\end{eqnarray*}
The goodness-of-fit analysis of the structural equation model above came to
contradictory indexes. \ac{CFI} and \ac{TLI} were relatively high (0.921 and
0.913, respectively) yet \ac{RMSEA} was not very small (0.114) and its p-value
was approximately zero, thus suggesting that the close fit hypothesis should be
rejected.  Besides, only the latent variable ``user'' was a significant latent
variable explaining relevance (i.e. qrel).  The regression coefficient was
indeed 0.230 (p-value was approximately zero), thus suggesting that the number
of clicks was a good predictor of relevance and that content, graph, link and
page were little significant in explaining relevance.  As for the relationships
between latent variables and manifest variables, the variables based on ``qtn'',
\ac{TFIDF} and \ac{LM} were the most significant in explaining content, inlink
was the most significant in explaining graph, the variables based on ``anchor''
in explaining url, quality in explaining page, and url\_clickcount was the most
significant manifest variable in explaining user.

\ac{SEM} cannot suggest true conceptual relationships nor can it tell whether a
variable can be the cause of another variable.  As regards the
eliteness-frequency-relevance relationship, in particular, \ac{SEM} cannot
suggest the true structural equation model; it can only tell whether the
observed data fit the given structural equation model.  It follows that a
structural equation model might not be the only model that fit the observed data
and that an alternative model may fit the data as well.  For example, the
following structural equation model
\begin{eqnarray}
  \label{eq:eliteness-3}
  \mbox{qrel} \rightarrow \mbox{eliteness} \rightarrow \log \mbox{bm25}+\mbox{doclen}
\end{eqnarray}
is another good fit of the observed data.  The problem with the
\eqref{eq:eliteness-3} is that an \ac{IR} researcher would perceive eliteness as
little likely related to doclen although the regression coefficient is
significantly different from zero and the approximate fit indexes suggest that
the model is a good fit.  In general, adding variables does not always decrease
the fit, but it makes a model less readable than another model that includes
fewer variables.

Latent variable names cannot be provided by \ac{SEM} which leaves a great deal
of freedom to the researcher who might, for example, replace ``eliteness'' with
``aboutness'' and obtain the same good approximate fit of the same structural
equation model, which first explains aboutness by relevance and then explains
bm25 and doclen by aboutness.  The naming just mentioned is an example of naming
fallacy.  The name of a latent variable cannot be considered a sufficient
condition that the latent variable is correct.  However, latent variables have
to be named to make them explicit to other researchers and in general to
readers.  In other domains, the issue is the same; for example, designers of
database conceptual schemas name entities and relationships and report on their
meanings by means of glossaries; moreover, data miners test clustering algorithms
that yield clusters that should be named and described to convey the nature of
the cluster points.  The semantics of a structural equation model does not only
depend on names, but it is also given by the complex of variables and associations.
Besides, the seeming limitation can be surpassed by explicitly reporting the
meaning of the names used in a structural equation model.

Another issue of \ac{SEM} is that a latent variable might not correspond to an
entity conceived by everyone in only one way; for example, eliteness might be
conceived as a small subset of terms by a researcher, whereas it might be
conceived as a more complex entity by another researcher.  Since latent variable
names are usually nouns, they suffer from the usual natural language drawbacks;
for example, a latent variable name may be a synonym of another name or may be
polysemous and carry more than one meaning at the same time.

Authoritativeness and aboutness are unrelated as shown by
\eqref{eq:auth-about-3}.  This outcome confirms the early literature on the use
of link analysis in \ac{IR} in that authoritativeness and aboutness should be
considered as distinct dimensions of relevance, capturing different user's
information needs; some users may require authoritative documents which might be
little relevant while other users may require relevant documents yet little
authoritative.  The lack of relationship between authoritativeness and aboutness
can also be observed by the lack of significance of the regression coefficient
of qrel in \eqref{eq:auth-about-2} as opposed to the significance of qrel in
\eqref{eq:auth-about-1}, thus suggesting that \ac{qrel} can be a signal of
aboutness and not of authoritativeness.

Indegree and pagerank are both significant manifestations of authoritativeness.
One reason for this simultaneous, significant manifestation may be due to the
relationship between pagerank and indegree.  Although it is a more complex
algorithm than simply counting in-links, PageRank and indegree are strongly
correlated (Pearson's product-moment correlation comes out to be $0.832$ with
$\mbox{p-value} \approx 0$).  To check the hypothesis that pagerank might be
removed from \eqref{eq:auth-about-1}, the fit of the structural equation model
introduced above has been recalculated without pagerank, thus obtaining very
similar results: exact fit test passed with $\mbox{p-value} = 0.570$;
approximate fit test passed with $\mbox{p-value} = 0.662$; $\mbox{\ac{CFI}} =
1$.

The structural equation model that relates relevance degree (i.e. qrel) to five
latent variables (i.e. content, user, link, graph, and quality) was only
partially satisfactory, at least as far as query 22636 of \ac{MSLR} is
concerned.  The unsatisfactory fit of this structural equation model suggests
that the latent variables causing \ac{qrel} might be less straightforward than
those encoded by content, user, link, graph, and page, the latter often being
utilised to model contextual search according to \citeN{Melucci12b},
\citeN{OBrien&13} and \citeN{Park&14}.  For example, a manifest variable should
be related to more than one latent variable in an improved structural equation
model; however, the addition of relationships between variables might make a
model unidentifiable.

To overcome the limitations on the generality of the results caused by the
utilisation of one query, automated tools that perform such an analysis for many
queries and datasets should be designed and implemented.  As regards the
goodness-of-fit of the structural equation model, although the approximate close
indexes (\ac{CFI} and \ac{TLI}) were relatively high, other statistics suggested
that better models should be found (for example \ac{RMSEA} was not very small).
Unfortunately, \ac{SEM} cannot straightforwardly suggest the correct and best
model unless the researcher helps to find such a model by using his knowledge
of the application domain, yet some help can be given by stepwise regression.

\subsection{Effect of Query Terms}
\label{sec:effect-query-terms}

In this section, the impact of the query term weights of Lucene's implementation
of the \ac{VSM}-based retrieval function and that of the \ac{BM25}-based
retrieval function will be investigated.  The \ac{VSM}-based retrieval function
is a modification of the classical \ac{VSM} retrieval function and was applied
for each query $Q$ and document $D$ as follows: \begin{eqnarray}
  \label{eq:vsm-lucene}
  \sum_{t \in Q} \mbox{dtw}_{t,D} \mbox{qtw}_{t,Q} \mbox{coord}_{Q,D} \mbox{boost}_t 
\end{eqnarray}
where 
\begin{eqnarray*}
  \mbox{dtw}_{t,D} &=& \frac{\mbox{tfidf}_{t,D}}{\mbox{length}_D} 
  \\ 
  \mbox{qtw}_{t,Q} &=& \frac{\mbox{tfidf}_{t,Q}}{\mbox{length}_Q}
  \\
  \mbox{coord}_{Q,D} &=& \frac{|D \cap Q|}{|Q|}
  \\
  \mbox{boost}_t &=& 1
\end{eqnarray*}
On the other hand, the \ac{BM25}-based run was obtained by the following
retrieval function
\begin{eqnarray}
  \label{eq:bm25-lucene}
  \sum_{t \in Q} \mbox{idf}_{t} \mbox{sat}_{t,D}
\end{eqnarray}
where 
\begin{eqnarray*}
  \mbox{idf}_{t,D} &=& \log\frac{N-\mbox{df}_t+0.5}{\mbox{df}_t+0.5} 
  \\ 
  \mbox{sat}_t &=& \frac{\mbox{tf}_{t,D}}{K + \mbox{tf}_{t,D}}
  \\
  K &=& k_1\left(1-b + b\frac{\mbox{doclen}}{\mbox{avdoclen}}\right)
\end{eqnarray*}
For each query, two lists of documents were created -- one list for
each retrieval function.  Each retrieved document has been associated
to the assessment of relevance to the query and was joined to the
components of the weight function of each query term.  In particular,
each document retrieved by the \ac{VSM}-based retrieval function was
joined to $\mbox{dtw}_{t,D}, \mbox{qtw}_{t,Q}, \mbox{coord}_{Q,D},
\mbox{boost}_t$ for each query term $t$, and each document retrieved
by the \ac{BM25}-based retrieval function was joined to
$\mbox{idf}_{t} \mbox{sat}_{t,D}$ for each query term $t$.  Moreover,
\ac{P@r} was computed for each document retrieved at rank $r$.

The following structural equation model was estimated as for the
\ac{BM25}-based run and query 305 (``Most Dangerous Vehicles: Which are
the most crashworthy, and least crashworthy, passenger vehicles?''):
\begin{eqnarray*}
  \mbox{prec} &\gets& \mbox{bm25}_{\mbox{\scriptsize
      crashworthy}}+\mbox{bm25}_{\mbox{\scriptsize dangerous}}+\\ &&\mbox{bm25}_{\mbox{\scriptsize passenger}}+\mbox{bm25}_{\mbox{\scriptsize vehicles}}
\end{eqnarray*}
The regression coefficients are as follows \footnote{The coefficients
  of ``crashworthy'' are not reported because only one retrieved
  document was indexed by ``crashworthy''.}:
\begin{eqnarray*}
  B_{\mbox{\scriptsize bm25,dangerous}} &=& 0.016 \\
  B_{\mbox{\scriptsize bm25,passenger}} &=& 0.023 \\ 
  B_{\mbox{\scriptsize bm25,vehicles}} &=& 0.017
\end{eqnarray*}
As the p-values were approximately zero, the regression coefficients
were significant; however, the fit was rather bad because of the low
number of exogenous variables. Although the regression coefficients
can be an interesting measure of the variation of prec, the beta
coefficients were of greater interest because they provide a measure
of the importance of each variable controlling the other variables.
In particular, the beta coefficients were:
\begin{eqnarray*}
  \beta_{\mbox{\scriptsize bm25,dangerous}} &=& 0.229 \\
  \beta_{\mbox{\scriptsize bm25,passenger}} &=& 0.397 \\ 
  \beta_{\mbox{\scriptsize bm25,vehicles}} &=& 0.342
\end{eqnarray*}
The beta coefficients are larger than the corresponding regression
coefficients because of the negative correlations between some pairs
of \ac{BM25} weights.  To investigate this model further, the
following structural equation model that replaces the \ac{BM25}
weights with their components (i.e. idf and sat) was estimated:
\begin{eqnarray*}
  \mbox{prec} &\gets& 
  \mbox{idf}_{\mbox{\scriptsize dangerous}}+
  \mbox{sat}_{\mbox{\scriptsize dangerous}}+
  \mbox{idf}_{\mbox{\scriptsize passenger}}+\\ &&
  \mbox{sat}_{\mbox{\scriptsize passenger}}+
  \mbox{idf}_{\mbox{\scriptsize vehicles}}+
  \mbox{sat}_{\mbox{\scriptsize vehicles}}
\end{eqnarray*}
The regression coefficients are as follows:
\begin{eqnarray*}
  B_{\mbox{\scriptsize idf,dangerous}} &=& 0.002 \\
  B_{\mbox{\scriptsize idf,passenger}} &=& 0.023 \\ 
  B_{\mbox{\scriptsize idf,vehicles}} &=& 0.018
\end{eqnarray*}
\begin{eqnarray*}
  B_{\mbox{\scriptsize sat,dangerous}} &=& 0.047 \\
  B_{\mbox{\scriptsize sat,passenger}} &=& 0.102 \\ 
  B_{\mbox{\scriptsize sat,vehicles}} &=& 0.082
\end{eqnarray*}
Except for $B_{\mbox{\scriptsize idf,dangerous}}$, these coefficients
are significant.  The corresponding beta coefficients are as follows:
\begin{eqnarray*}
  \beta_{\mbox{\scriptsize idf,dangerous}} &=& -0.035 \\
  \beta_{\mbox{\scriptsize idf,passenger}} &=& -0.323 \\ 
  \beta_{\mbox{\scriptsize idf,vehicles}} &=& -0.340
\end{eqnarray*}
\begin{eqnarray*}
  \beta_{\mbox{\scriptsize sat,dangerous}} &=& 0.301 \\
  \beta_{\mbox{\scriptsize sat,passenger}} &=& 0.813 \\ 
  \beta_{\mbox{\scriptsize sat,vehicles}} &=& 0.857
\end{eqnarray*}
The following structural equation model was estimated as for the
\ac{VSM}-based run:
\begin{eqnarray*}
  \mbox{prec} &\gets& \mbox{dtw}_{\mbox{\scriptsize crashworthy}} +
  \mbox{dtw}_{\mbox{\scriptsize dangerous}} + \mbox{dtw}_{\mbox{\scriptsize passenger}} +
  \mbox{dtw}_{\mbox{\scriptsize vehicles}} + \\ && \mbox{qtw}_{\mbox{\scriptsize crashworthy}} + \mbox{qtw}_{\mbox{\scriptsize dangerous}} + \mbox{qtw}_{\mbox{\scriptsize passenger}} + \mbox{qtw}_{\mbox{\scriptsize vehicles}}
\end{eqnarray*}
The regression coefficients are as follows and all were significant:
\begin{eqnarray*}
  B_{\mbox{\scriptsize dtw,dangerous}} &=& 0.064 \\ 
  B_{\mbox{\scriptsize dtw,passenger}} &=& 0.095 \\ 
  B_{\mbox{\scriptsize dtw,vehicles}} &=& 0.078
\end{eqnarray*}
\begin{eqnarray*}
  B_{\mbox{\scriptsize qtw,dangerous}} &=& 0.088 \\ 
  B_{\mbox{\scriptsize qtw,passenger}} &=& 0.042 \\ 
  B_{\mbox{\scriptsize qtw,vehicles}} &=& 0.074
\end{eqnarray*}
The beta coefficients were as follows:
\begin{eqnarray*}
  \beta_{\mbox{\scriptsize dtw,dangerous}} &=& 0.341 \\ 
  \beta_{\mbox{\scriptsize dtw,passenger}} &=& 0.929 \\ 
  \beta_{\mbox{\scriptsize dtw,vehicles}} &=& 0.635
\end{eqnarray*}
\begin{eqnarray*}
  \beta_{\mbox{\scriptsize qtw,dangerous}} &=& 0.412 \\ 
  \beta_{\mbox{\scriptsize qtw,passenger}} &=& 0.205 \\ 
  \beta_{\mbox{\scriptsize qtw,vehicles}} &=& 0.331
\end{eqnarray*}

The badness of fit of the structural equation models above depends on the low
number of exogenous variables; the fit was very good when all the weight
components were added to the model for each query term. A good fit may be useful
for prediction purposes; however, it may be misleading when the role played by
the \ac{BM25}-based query term weights is of interest.  It might be misleading
because, if all the weight components were added to the model for each query
term, the beta coefficients of the saturation weights would have the opposite
sign of the \ac{BM25}-based query term weights.  The difference in sign between
the beta coefficients of saturation and those of \ac{BM25} is counter-intuitive
since both saturation and \ac{BM25} should be positively correlated to
prec. However, the difference in sign is caused by the strong correlations
between the weight components that make the beta coefficients negative.  The
high collinearity could be acceptable when the variables are due to natural
processes such as the collinearity between height and weight. In the event, they
are not caused by natural processes; on the contrary, they are caused by the
mathematical formulation of the function which make \ac{BM25} functionally
dependent on saturation and IDF. It follows that the exogenous variables of the
weight components (e.g. IDF) should be ignored and not added to the structural
equation model together with \ac{BM25} in the analysis.

The negative correlation between \ac{BM25} weights can be quite surprising since
it is expected that all the query terms participate in increasing \ac{P@r}.
Instead, the results of the analysis suggest that when one query term
contributes to retrieval effectiveness, another query term is detrimental -- for
query 305 and the \ac{BM25}-based retrieval function at least.  The beta
coefficients of the four query terms above indicate the most important query
terms as regards \ac{P@r} when document ranking is performed by the
\ac{BM25}-based retrieval function.  When the retrieval functions are compared,
the regression coefficients have to be used.



\section{Conclusions and Future Directions}
\label{sec:conclusions}

The crucial point of the use of \ac{SEM} in data analytics is the definition of
the structural equation models that describe the observed data at their best. It
would be desirable to always find the best structural equation model, that is
the model that fit the data very well on the basis of statistically significant
parameters and of a reasonable narrative -- from the researcher's perspective at
least. However, the best model cannot always be found, since two or more models
may fit the observed data well or no fitting model may be found at all. Another
weakness is the need to define structural equation models (e.g. path models)
starting from  many manifest variables. Although the researcher's judgment
should always be considered, manually finding the best model requires a
considerable intellectual effort and some automatic method -- semi-automatic at
least -- would be desirable.

In the area of learning-to-rank, in particular, and in that of Machine Learning,
in general, a number of procedures for selecting features and fitting functions
have been developed \cite{Liu11}.  Although these procedures should be
considered with reference to the problem of defining and estimating structural
equation models, the selection of the variables of a structural equation model
is a more complex task than the definition of real functions of the scores and
weights which are observed for documents and terms to the aims of learning to
rank.  Variable selection has to do with the description of the retrieval models
such as the \ac{VSM}, the language models and the probabilistic models; the
question is how to represent a retrieval model in terms of variables,
relationships and therefore in terms of a structural equation model.

Moreover, further research would be advisable to find methods that ``translate''
a structural equation model into rules of modification for a more effective
retrieval model once the structural equation model has been found for the
retrieval model.  Indeed, the ultimate goal of the use of \ac{SEM} in \ac{IR}
evaluation would be the transformation of a retrieval model into a \emph{new},
more effective model.  Such a transformation resembles what the approaches to
learning-to-rank aim for, that is, a set of parameters of a real function
mapping an independent multi-variate variable to a dependent univariate
variable.

The potential of \ac{SEM} is the capacity to combine latent variables with
manifest variables. The ability of using latent variables that may be developed
may lead to implementing some general hypotheses about \ac{IR} (e.g.  the role
played by authoritativeness or search task) and their influence on retrieval
effectiveness. This ability may have some desirable effects. On the one hand, it
may facilitate the investigation of the processes of information seeking based
on the quantitative analysis provided by \ac{SEM}. On the other hand, it may
help the researchers to explain the results gathered throughout the course of
their experiments by using more effective statistical instruments than
descriptive or inferential statistics.

One distinguishing feature of \ac{SEM} is the graphical nature of a structural
equation model; such a model can be communicated in spoken or written words
because variables and causal relationships thereof may be viewed as concepts
(e.g. nouns) and associations (e.g. verbs).  As a result of the graphical nature
of a structural equation model, \ac{SEM} may become a new language helping the
researchers in \ac{IR} to find more powerful descriptions and explanations of
theoretical models and experimental results than traditional statistics. 

Despite the potential expressed since Wright's pioneering work
\citeyear{Wright&18}, some misunderstandings are still limiting the potential of
\ac{SEM} \cite{Bollen&13}. First, it is often believed that correlation implies
causation and that a significant regression coefficient may be considered a
strong signal that a variable is the cause of another variable.  Instead,
\ac{SEM} cannot discover causal relationships other than the relationships
already encoded in the researcher's structural equation model.  Second, \ac{SEM}
is often viewed as nothing but a complicated regression and \ac{ANOVA}
technique.  Causal networks rather, allow the researchers to utilise a language
that is not part of standard statistics for expressing their application domains
differently from the way provided by regression and \ac{ANOVA}
\cite{Pearl&09a,Pearl12a}.


\newpage

\addcontentsline{toc}{section}{Bibliography}
\end{document}